\documentclass{ws-ijmpa}
\usepackage[super,compress]{cite}
\usepackage{graphicx}
\usepackage{color}

\begin{document}

\markboth{C. A. Escobar and L. F. Urrutia}
{Extended Nambu Models: their relation to gauge theories}

%%%%%%%%%%%%%%%%%%%%% Publisher's Area please ignore %%%%%%%%%%%%%%%
%
\catchline{}{}{}{}{}
%
%%%%%%%%%%%%%%%%%%%%%%%%%%%%%%%%%%%%%%%%%%%%%%%%%%%%%%%%%%%%%%%%%%%%

\title{Extended Nambu Models: their relation to gauge theories}

\author{C. A. Escobar}
\address{CENTRA, Departamento de F\'{i}sica, Universidade do Algarve, 8005-139 Faro, Portugal. \\ cruiz@ualg.pt}

\author{L. F. Urrutia}
\address{Instituto de Ciencias Nucleares, Universidad Nacional Aut\'{o}noma de M\'{e}xico, 04510 M\'{e}xico, Distrito Federal, M\'{e}xico \\  urrutia@nucleares.unam.mx}

\maketitle

\begin{history}
\received{Day Month Year}
\revised{Day Month Year}
\end{history}

\begin{abstract}
Yang-Mills theories supplemented by an additional coordinate constraint, which is solved and substituted in the original Lagrangian, provide examples of the so called Nambu models, in the case where such constraints arise from spontaneous Lorentz symmetry breaking. Some explicit calculations have shown that, after additional conditions are imposed, Nambu models are capable of reproducing the original gauge theories, thus making Lorentz violation unobservable and allowing the interpretation of the corresponding massless gauge bosons as  the Goldstone bosons arising from the spontaneous symmetry breaking. A natural question posed by this approach in the realm of gauge theories is to determine under which conditions the recovery of an arbitrary gauge theory from the corresponding  Nambu model, defined by a general constraint over the coordinates, becomes possible. We  refer to these theories as extended Nambu models (ENM) and emphasize the fact that  the {defining coordinate} constraint is not treated as a standard gauge fixing term. At this level, the mechanism for generating the constraint is irrelevant and the case of spontaneous Lorentz symmetry breaking is taken only as a motivation, which naturally  bring  this problem under consideration. Using a non-perturbative Hamiltonian analysis we prove that the ENM yields the original gauge theory  after we demand current conservation for all time, together with  the imposition of the Gauss laws constraints as initial conditions upon the dynamics of the ENM. The Nambu models yielding  electrodynamics, Yang Mills theories and linearized gravity are particular examples of our general approach.
\end{abstract}

\ccode{PACS numbers: 11.15.-q, 11.15.Ex, 14.70.-e, 11.10.Ef}

\section{Introduction}

Gauge theories, symmetry principles and spontaneous symmetry
breaking have been successfully put together in building the electroweak sector of the
Standard Model \cite{Electrodebil,Electrodebil2}. Another well known example including
these concepts is the construction of pion interactions in the nonlinear
sigma model, which is characterized by spontaneous chiral symmetry breaking 
\cite{chiral,chiral2,chiral3}. Here, pions are interpreted as the massless Goldstone bosons
(GBs) generated by such breaking and the construction of their interactions
arises from the most general Lagrangian respecting the remaining unbroken
symmetries. It was precisely the understanding of pions as GBs what
motivated the possibility of looking at fundamental massless particles, like
photons and gravitons for example, as the GBs arising from some spontaneous
symmetry breaking. Since these particles are of tensorial nature, a tensor-valued vacuum expectation value 
(VEV) would be required. Such a non-zero VEV 
produces fixed directions on the spacetime, thus leading to
spontaneous Lorentz symmetry breaking (SLSB). The main goal of this proposal is to provide a
dynamical setting for the gauge principle. One of the first realizations of this idea is the abelian
Nambu model (ANM), which was proposed in Ref. \citen{Nambu-Progr} as a description of
electrodynamics arising from SLSB. The ANM is defined by the standard
Maxwell Lagrangian plus a constraint over the vector potential 
\begin{equation}
\mathcal{L}(A_{\mu })=-\frac{1}{4}F_{\mu \nu }F^{\mu \nu }-A_{\mu }J^{\mu
},\quad \quad \quad A_{\mu }A^{\mu }=\mathfrak{n}^{2}M^{2},
\label{NANMabeliano}
\end{equation}
where $\mathfrak{n}^{\mu }$ is a properly oriented constant vector in the
Lorentz space, while $M$ is the proposed scale associated with the SLSB. The
origin of such a constraint can be understood as the consequence of a non-zero
vacuum expectation value of the vector potential: $\langle A_{\mu }\rangle =
\mathfrak{n}_{\mu }M$ produced by the SLSB. In a way similar to the non linear sigma model, the constraint is to be solved and substituted in the Lagrangian ${\mathcal L}$, thus drastically modifying the properties of the original gauge theory. In fact, the resulting model defined by Eq. (\ref
{NANMabeliano})   has more degrees
of freedom (DOF),  current conservation is not fulfilled
 and it does not exhibit gauge invariance. The aim in Ref. \citen{Nambu-Progr} was to make explicit the
conditions under which the ANM turns out to be  equivalent to standard QED,
instead of yielding a physical violation of the Lorentz symmetry.
The idea that gauge particles (photons and gravitons, for example) might arise as
the GBs of a theory with SLSB has been widely studied and goes back a long
way\cite{RUSOS,Bjorken,Guralnik,Guralnik1}. In this approach, the masslessness of these
gauge particles can then be understood in terms of the Goldstone theorem 
\cite{TeoGoldstone,TeoGoldstone2}, instead of gauge invariance requirements. Nambu models have been further considered
in relation to electrodynamics \cite{Azatov-Chkareuli,Urru-Mont} and
generalized to the Yang Mills \cite{JLCH1,JLCH2,JLCH3,NANM_ES_UR} and the
gravitational \cite{JLCH4} cases. Perturbative calculations in  Nambu models show that, to the order
considered and under some appropriate initial conditions, all SLSB
contributions to physical processes cancel out, yielding an equivalence with
the original gauge theory \cite{Nambu-Progr,Azatov-Chkareuli,JLCH1,JLCH2,JLCH3,JLCH4}. Additional previous works in gravitation can be found in Refs. \citen{CARROLL,Kraus,Kosteleckygravity,Kosteleckygravity2}, respectively.

In this way, Nambu models can be understood as a gauge theory plus some constraint upon the coordinates, which is to be solved and substituted in the corresponding Lagrangian, thus producing a quite different model as a first step. However, since the final goal is to recover the original gauge theory, this motivates the problem of determining under which conditions this is accomplished. This non-standard procedure is illustrated for a Yang-Mills  theory in Ref. \citen{NANM_ES_UR}.
In this paper we deal with this question for what we call an extended Nambu model (ENM) and emphasize the fact that  the constraint is not treated as a standard gauge fixing term. The ENM is  defined by an  arbitrary gauge invariant
Lagrangian density, which we call the mother
gauge theory (MGT), supplemented by a constraint among the coordinates analogous to that appearing in Eq. (\ref{NANMabeliano}). From the general point of view we adopt here, the way in which the constraint is generated is irrelevant for our purposes and the case  of SLSB is taken only as a motivation, which naturally  bring  this problem under consideration. The MGT is defined by a Lagrangian density $\mathcal{L}$ from which only  first class
constraints (FCC) arise, being the generators of a non trivial gauge symmetry. 

The paper is organized as follows. In Section \ref{gaugetheory} we define
the general MGT which we are going to deal with. Then we employ the Dirac
method \cite{Dirac_method} to perform the Hamiltonian analysis leading to
presence of FCC. We close this section by writing the Hamiltonian and the
canonical algebra that define the dynamics of the MGT, which will be used as
benchmark to establish the equivalence between this MGT and the
corresponding ENM. In Section \ref{NGmodel} we present the ENM, which is
defined by the same Lagrangian density $\mathcal{L}$ introduced in Section 
\ref{gaugetheory}, plus one constraint $F$ among the coordinates of the MGT. This constraint is
solved for two generic cases and subsequently substituted in the MGT Lagrangian.
Such an ENM leads to a theory without gauge invariance, which is manifest in
the appearance of second class constraints only. Under the conditions
imposed for $\mathcal{L}$ in Section \ref{gaugetheory}, we subsequently
perform the Hamiltonian analysis of the ENM. Finally, identifying a
suitable transformation between the canonical variables of the MGT and those
of the ENM, we show that the Hamiltonian describing the MGT and the
Hamiltonian corresponding to the ENM have the same functional form. Also,
such transformation allows us to prove that the canonical algebra of the ENM
induces the canonical algebra of the MGT. In this way, after suitable 
conditions are imposed in order to recover gauge invariance, the ENM
is shown to be equivalent to the original MGT. In Section \ref{examples}, we
present some examples of MGT where the conditions required in Section \ref
{gaugetheory} are fulfilled and, as a consequence, the equivalence with any
associated ENM, defined by the constraint $F$, is established. The \ref{CURCONS1} includes the calculation  of the generalized current conservation equation in our MGT, which arises as a consequence of the gauge invariance of the Hamiltonian action. \ref{Mechanics1a} contains a detailed discussion  of the method employed in the case of a simple mechanical model. In \ref{Apendice1} and \ref{Apendice2}, we explicitly show that the transformations between the canonical
variables of the MGT and those of the ENM yield the canonical algebra of the
former starting from the canonical algebra of the latter. To this end we
also need to calculate the algebra of the second class constraints of the
ENM, which is relevant to establish the relation between the canonical
algebras of both theories.

\section{Mother Gauge Theories}

\label{gaugetheory} 

In this section we state the conditions that define the
class of MGT that we are going to deal with. We will show that these
conditions allow for the construction of an ENM, which ultimately turns out
to be equivalent to the original MGT, after some conditions are
imposed. We will prove this by construction.

The MGT is defined by the Lagrangian density 
\begin{equation}
\mathcal{L}=\mathcal{L}(\theta_{l},\lambda _{A},\dot{\theta}_{l}),\quad \quad \quad
\quad l=1,2,...,n.\quad \quad \quad A=1,2,...,K.  \label{Lagrangian}
\end{equation}
where both $\theta_{l}$ and $\lambda _{A}$ are independent field coordinates  depending upon the spacetime labels $t, {\mathbf x}=\{x_a; a=1,2,3\}$. To maintain the notation simple, in the following we do  not write the spacetime dependence of the fields, unless some confusion arises. Here $
\mathcal{L}$ is independent of the velocities $\dot{\lambda}_{A}$. 
To consider the coupling with external currents, we can split the Lagrangian density as
\begin{equation}
\mathcal{L}= \mathcal{L}_0(\theta_{l},\lambda _{A},\dot{\theta}_{l})-(\theta_lJ_l+\lambda_A J_A).
\end{equation}
The term $-(\theta_lJ_l+\lambda_A J_A)$ is a standard way to couple an external current, which  looks like  {$-{\rm Tr}(\mathbf{A}_\mu \mathbf{J}^\mu$)} in the covariant form of the  Yang Mills case. 
We are going to deal with theories where a generalized form of current conservation is satisfied and we will require to  establish how this property is expressed in our {case}.
Notice that the functional form of $\mathcal{L}$ is completely arbitrary. The
coordinates $\lambda _{A}$ play a similar r\^ole to the standard Lagrange
multipliers in a gauge theory and ensure that primary constrains appear. 

As usual, the canonical momenta are given by 
\begin{equation}
\Pi _{l}^{\theta}=\frac{\partial \mathcal{L}}{\partial \dot{\theta}_{l}},\quad \quad
\quad \quad \Pi _{A}^{\lambda }=\frac{\partial \mathcal{L}}{\partial \dot{
\lambda}_{A}},  \label{moments}
\end{equation}
with the {equal time} non-zero canonical Poisson brackets (PBs) algebra 
\begin{equation}
\{\theta_{l}(\mathbf{x}),\Pi _{k}^{\theta}(\mathbf{y})\}=\delta _{lk} \delta^3(\mathbf{x}-\mathbf{y})\,,\quad \quad \{\lambda _{A}(\mathbf{x}),\Pi
_{B}^{\lambda}(\mathbf{y})\}=\delta _{AB} \delta^3(\mathbf{x}-\mathbf{y}).  \label{commutationfree}
\end{equation}
In general we will dispense of the coordinates dependence of the PB's and we will write 
$\{\lambda _{A},\Pi
_{B}\}=\delta _{AB}$, for example. This encodes the implicit assumption that the corresponding subindexes also carry the space coordinate dependence.

{In
order to have a familiar perspective of the procedure, we will keep in mind
the {four dimensional} Yang Mills case as a guiding example. Here we identify {$\theta_{l}\rightarrow
A_{a}^{\alpha}$,  $\lambda _{A}\rightarrow A_{0}^{\alpha}$, $\Pi_l^\theta \rightarrow \Pi^\alpha_a=-E^\alpha_a$, $\Pi^\lambda_A \rightarrow \Pi^\alpha_0$, $J_l \rightarrow J^\alpha_a$ and $J_A \rightarrow J_0^\alpha$  where $\alpha$ and $
(0,a=1,2,3)$ }denote the indices in the gauge group and in the Lorentz space,
respectively.}

Given that the velocities $\dot{\lambda}_{A}$ are not present in $\mathcal{L}
$, the theory has primary constraints $\Phi _{A}^{1}=\Pi _{A}^{\lambda }=0$
and we take the standard Dirac method to perform the Hamiltonian analysis,
starting from the canonical Hamiltonian density 
\begin{equation}
\mathcal{H}_{c}=\dot{\theta}_{l}\Pi _{l}^{\theta}-\mathcal{L}.
\end{equation}
As a matter of notation, for any label $Z$, we will always write $H_Z$ for the Hamiltonian corresponding to the Hamiltonian density ${\mathcal H}_Z(y)$, i. e. $H_Z=\int d^3y {\mathcal H}_Z(y) $.

From now on, we restrict ourselves to MGTs satisfying the next three
conditions:
\begin{enumerate}
\item The canonical Hamiltonian density is linear in the coordinates $
\lambda _{A}$ and can be written as follows 
\begin{equation}
\mathcal{H}_{c}=\mathcal{H}_{F}(\theta_{i},\Pi _{i}^{\theta})+\lambda
_{A}(G^0_{A}(\theta_{i},\Pi _{i}^{\theta})+ J_A) +\theta_l J_l.  \label{hamiltonianocanonic}
\end{equation}
{The corresponding Hamiltonian is $H_c$}. We refer to the quantities 
{\begin{equation}
G_{A}(\theta_{i},\Pi _{i}^{\theta})\equiv G^0_{A}(\theta_{i},\Pi _{i}^{\theta}) +J_A
\end{equation} }
as Gauss functions. They will turn out to be FCC in the MGT (the corresponding Gauss laws), but not in the ENM.
Let us emphasize that $J_A$ and $J_l$, being external currents, do not play any r\^ole in the calculation
of PBs.
\item The only primary constraints are 
\begin{equation}
\Phi _{A}^{1}=\Pi _{A}^{\lambda }=0,  \label{condition2}
\end{equation}
i.e., the momenta canonically conjugated to the coordinates $\lambda_{A}$.
\item The conditions 
\begin{eqnarray}
&&\{G^0_{A}(\mathbf{x}),\mathcal{H}_{F}(\mathbf{y}\}=C_{AB}G^0_{B}(\mathbf{x})\delta^3(\mathbf{x}-\mathbf{y}),\quad \quad \nonumber \\
&&\{G^0_{A}(\mathbf{x}),G^0_{B}(\mathbf{y})\}=C_{ABC}G^0_{C}(\mathbf{x}) \delta^3(\mathbf{x}-\mathbf{y}),  \label{condition3}
\end{eqnarray}
hold at equal times.
\end{enumerate}
For example, in the Yang Mills case we have {$G_{A}(\theta_{i},\Pi _{i}^{\theta})\rightarrow -(D_aE_a-J_0)^\alpha$} with {$G^0_A\rightarrow -(D_aE_a)^\alpha$}. {Here $D_a$ denotes the corresponding covariant derivative.}

We shall prove that the conditions (\ref{hamiltonianocanonic}), (\ref
{condition2}) and (\ref{condition3}) are enough to ensure that we are
dealing with a theory having only first class constraints. We also demand that the  constraints $G_A$  generate non-trivial gauge symmetry transformations, which must be   verified in each particular case. In this way, such transformations will lead to the gauge invariance of  the action.  We assume a positive answer in the following. At the same time, a full Hamiltonian
analysis, respecting such conditions, will be done in order to determine:
the number of degrees of freedom, the identification of the  constraints, the extended Hamiltonian density and the canonical
algebra.

Following the Dirac method, the extended Hamiltonian density is 
\begin{equation}
\mathcal{H}_{E}=\mathcal{H}_{c}+\beta _{A}\Pi _{A}^{\lambda }\;=\mathcal{H}
_{F}+\lambda _{A}G_{A}+\theta_lJ_l+\beta _{A}\Pi _{A}^{\lambda },
\end{equation}
where $\beta _{A}$ are arbitrary functions. {The corresponding Hamiltonian is $H_E$}. The time evolution condition of
the primary constraints $\Phi _{A}^{1}=\Pi _{A}^{\lambda }$ yields 
\begin{equation}
\dot{\Phi}_{A}^{1}(\mathbf{x})=\{\Pi _{A}^{\lambda }(\mathbf{x}), H_E \}=\int d^{3}y\,\{\Pi _{A}^{\lambda }(
\mathbf{x}),\lambda _{B}(\mathbf{y})\}G_{B}(\mathbf{y})=-G_{A}(\mathbf{x}),
\end{equation}
i.e, to the secondary constraints $\Phi _{A}^{2}=G_{A}$. Let us remark that the constraints turn out to be the  $G_{A}$'s instead of the $G^0_{A}$'s, in such a way that in some cases it is useful to express the results in the right hand side of Eq.(\ref{condition3}) in terms of the former. Given that $
G_{A}=G_{A}(\theta_{i},\Pi _{i}^{\theta})$, we have $\{\lambda _{B},G_{A}\}=0$ and $
\{\Pi _{B}^{\lambda },G_{A}\}=0$. Considering Eq. (\ref{condition3}), it
follows that the time evolution condition of the secondary constraints $\dot{
\Phi}_{A}^{2}$ is 
\begin{equation}
\dot{\Phi}_{A}^{2}=\dot{G}_A=\{G_{A},{H}_{E}\}\approx \dot{J}_A+ C_{AB}G^0_{B} +{C}_{ABC}\lambda_B G^0_{C}+\int d^3y \, \{G^0_{A},\theta_l(\mathbf y)\}J_l(\mathbf{y}),
\label{gaussbracket}
\end{equation}
Introducing the constraints $G_A \approx 0$ in the above equation we can write $\dot{G}_A \approx (DJ)_A$, where 
\begin{equation}
 (DJ)_A \equiv \dot{J}_A-C_{AB}J_B- C_{ABC}\lambda_BJ_C+\int d^3y \, \{G^0_{A},\theta_l(\mathbf y)\}J_l(\mathbf{y}). 
\label{DEFDJA}
\end{equation}
In \ref{CURCONS1} we show that
\begin{equation}
(DJ)_A=0,
\label{currentcons}
\end{equation}
as a consequence of the invariance of the Hamiltonian action under the gauge transformations generated by the first class constraints $G_A$, which correspond to the generalized Gauss laws. In this way  $(DJ)_A=0$ is the generalized version of current conservation in our model. Once the explicit form of the generators $G_A$ is given, one can recover the standard expressions in terms of the covariant derivatives, for example. A detailed calculation of the right hand side of Eq.(\ref{DEFDJA}) in the case of a Yang-Mills theory can be found in the Appendix A of Ref. \citen{NANM_ES_UR}.

Under the condition (\ref{currentcons}), the quantities $\dot{\Phi}_{A}^{2}$ are weakly equal zero and there
are no more constraints in the theory. We remark that the PBs among the quantities $(\Pi
_{A}^{\lambda },G_{B})$ are zero or weakly zero, leading to the appearance
of $2K$ first class constraints. The presence of these quantities implies
that we are dealing with a proper gauge theory.

We have $n+K$ variables in coordinate space and $
2K$ first class constraints, which implies that the number of degrees of
freedom (DOF) is 
\begin{equation}
\#\text{DOF}=\frac{1}{2}\bigg(2(n+K)-2(2K)\bigg)=n-K.  \label{dofgauge}
\end{equation}
For a Yang-Mills theory, when the number of generators of the gauge group is $N$, we have $n=3N$
and $K=N$, yielding the correct number of $2N$ DOF.

In general, the presence of first class constraints implies unphysical
degrees of freedom, which might be conveniently removed to obtain the
reduced phase space. To this end we have to fix the gauge by imposing as
many suitable gauge constraints as the number of first class constraints
that we want to eliminate. These gauge constraints have to be admissible
and should convert the set of gauge conditions plus the set of first class
constraints to be eliminated into a set of second class constraints.
Subsequently, each constraint in this set is set strongly equal to zero,
after the introduction of the corresponding Dirac brackets (DBs). In our case
we choose to eliminate the variables $\lambda _{A}$ and $\Pi _{B}^{\lambda}$. 
The gauge is partially fixed by adding to Eq. {(\ref{condition2})} the constraints 
{\begin{equation}
\Phi
_{A}^{3}=\lambda _{A}-\Theta _{A}\approx 0,
\end{equation} }
where $
\Theta _{A}$ are arbitrary functions to be consistently determined after the
remaining first class constraints $G_{A}$ are fixed. The constraints $\Phi
_{A}^{3}$ and $\Phi _{B}^{1}=$ $\Pi _{B}^{\lambda }$ become in fact second
class, i.e. the matrix $Q_{AB}=\{\Phi _{A}^{1},\Phi _{B}^{3}\}$ can be
inverted. Fixing strongly $\Phi _{A}^{1}=0$ and $\Phi _{B}^{3}$ $=0\;$we
obtain, after introducing the Dirac brackets, the partially reduced
Hamiltonian density 
\begin{equation}
\mathcal{H}_{E}=\mathcal{H}_{F}+\Theta _{A}G_{A}+\theta_lJ_l.  \label{hamiltonianofinal}
\end{equation}
In order to compute the corresponding Dirac brackets $\{\mathcal{A},\mathcal{
B}\}_{D}$, we require the $2N\times 2N$ matrix constructed with the PBs of the constraints $\Phi_A^1$ and $\Phi_B^3$
\begin{equation}
M=\left( 
\begin{array}{cc}
0 & Q \\ 
-Q^{T} & R
\end{array}
\right) ,
\label{MATRIX}
\end{equation}
where $Q=[Q_{AB}]$, $R=[R_{AB}]$ with $Q_{AB}=\{\Phi _{A}^{1},\Phi
_{B}^{3}\} $, $R_{AB}=\{\Phi _{A}^{3},\Phi _{B}^{3}\}$. Here we have made
use of the PBs $\{\Phi _{A}^{1},\Phi _{B}^{1}\}=0=\{\Pi _{A}^{\lambda },\Pi
_{B}^{\lambda }\}$. The inverse matrix is given by 
\begin{equation}
M^{-1}=\left( 
\begin{array}{cc}
(Q^{T})^{-1}RQ^{-1} & -(Q^{T})^{-1} \\ 
Q^{-1} & 0
\end{array}
\right) .  \label{inversa}
\end{equation}
Let us clarify another point in the notation. The matrix elements in Eq. (\ref{MATRIX}), for example, also carry  space-coordinate labels which are suppressed. In other words, $M_{AB}$ really stands for the equal-time object $M_{AB}(t, {\mathbf x};t,{\mathbf y})$. This is relevant  for matrix multiplication where the product $M_{AB}=P_{AC}Q_{CB}$ corresponds to $M_{AB}(t,{\mathbf x}; t,{\mathbf y})=\int d^3 z \, P_{AC}(t,{\mathbf x};t, {\mathbf z})Q_{CB}(t, {\mathbf z}; t,{\mathbf y})$. The matrix elements are evaluated at fixed time $t$, so that the notation can be further simplified  
to $M_{AB}(x,y)=\int d^3z \, P_{AC}(x,z)Q_{CB}(z,y)$.

The equal-time Dirac brackets are defined as 
\begin{equation}
\{\mathcal{A}(\mathbf{x}),\mathcal{B}(\mathbf{y})\}_{D}=\{\mathcal{A}(
\mathbf{x}),\mathcal{B}(\mathbf{y})\}-\int \,d^{3}u\,d^{3}v\{\mathcal{A}(
\mathbf{x}),\chi _{i}(\mathbf{u})\}(M^{-1})^{ij}\{\chi _{j}(\mathbf{v}),
\mathcal{B}(\mathbf{y})\},
\end{equation}
where $\chi _{j}$ denote any of the constraints $\Phi _{A}^{1}$ and $\Phi
_{B}^{3}$. Using $M^{-1}$ given in (\ref{inversa}), together with the fact
that $\{\Phi _{A}^{1},\theta_{j}\}=\{\Phi _{A}^{1},\Pi _{j}^{\theta}\}=0$ we obtain
the canonical algebra for the remaining variables $\theta_{i}$ and $\Pi _{j}^{\theta}$
\begin{equation}
\{\theta_{i},\theta_{j}\}_{D}=0,\quad \quad \quad \{\Pi^\theta _{i},\Pi
_{j}^{\theta}\}_{D}=0,\quad \quad \quad \{\theta_{i},\Pi _{j}^{\theta}\}_{D}=\delta _{ij}.
\label{algebrafinal}
\end{equation}
We could further fix the constraints $G_{A}$, however at this point the
Hamiltonian (\ref{hamiltonianofinal}) together with the canonical algebra (\ref{algebrafinal})
 are sufficient to determine the dynamics of the MGT.

\section{The Extended Nambu Model}

\label{NGmodel}

In this section we define the extended Nambu model (ENM) associated with the
MGT previously introduced. Our main goal is to establish the equivalence
between both models by finding which additional conditions have to be
imposed upon the ENM in order that its Hamiltonian and canonical algebra
reduce to (\ref{hamiltonianofinal}) and (\ref{algebrafinal}), respectively.
We will prove that the aforementioned equivalence can be reached
provided that the conditions (\ref{hamiltonianocanonic}), (\ref
{condition2}) and (\ref{condition3}) are fulfilled.

The ENM is given by the same Lagrangian (\ref{Lagrangian}) defining the MGT 
\begin{equation}
\mathcal{L}=\mathcal{L}(\theta_{l},\lambda _{A},\dot{\theta}_{l}),\quad \quad \quad
l=1,...,n.\quad \quad \quad A=1,...,K.  \label{lag2}
\end{equation}
plus one constraint
\begin{equation}
F(\theta_{l},\lambda _{A})=0.  \label{constriccion}
\end{equation}
In the standard abelian Nambu model, $\mathcal{L}=\mathcal{L}(A_{\mu })=
\mathcal{L}(A_{i},A_{0},\dot{A}_{i})$ corresponds to the Maxwell Lagrangian
density, $\theta_{i}\rightarrow A_{i}$, $\lambda \rightarrow A_{0}$ and $F(A_{\mu
})=A_{\mu }A^{\mu }-n^{2}M^{2}$. Here, we consider a wider class\ of
Lagrangian densities $\mathcal{L}$ and constraints $F$ to define the ENM. As in the {case of the} MGT, {where external currents are considered}, the Lagrange density and the quantities $G_A$ will be split into $\mathcal{L}=\mathcal{L}_0-(\lambda_AJ_A+\theta_lJ_l)$ and $G_A=G_A^0+J_A$, respectively.  

For the moment, we deal with the case where $F(\theta_{i},\lambda _{A})=0$ includes all the variables $\lambda _{A}$.  
The function $F=F(\theta_l,\lambda_A)$ is not completely arbitrary but must satisfy some general conditions to be
determined in the \ref{Apendice1} and {the} \ref{Apendice2}, depending how the constraint F is solved.  

In comparison with the MGT, the introduction  of the additional constraint {in Eq. (\ref{constriccion})}, which is not {handled} as a gauge fixing {condition}, drastically
modifies the structure and the dynamics of the {ENM}. In this way, it is
not completely straightforward how an equivalence between the ENM and the
MGT can be established. To mention just an example,  
 significant differences between the ANM and the standard electrodynamics are: (i) the ANM  has only second class constraints (there is no gauge
invariance), (ii) the number of DOF in the ANM is three, while standard
electrodynamics has only two DOF, (iii) the equations of motion do not
match and (iv) current conservation does not follow from the equations of motion  in the ANM.

The general procedure through which we analyze the ENM is by solving
explicitly the constraint (\ref{constriccion}) for one variable and
substituting this solution into the Lagrangian density (\ref{lag2}). There
are several ways to solve Eq. (\ref{constriccion}) and we present the two
generic cases, which reduce to solve either for one coordinate $\theta_l$ or for
one coordinate $\lambda_A$. Both cases yield the same conditions for the
equivalence we intend to establish.

The general strategy is: (1) After solving
the constraint (\ref{constriccion}) we identify the canonical variables of
the ENM, together with its canonical algebra. (2) Since both theories arise
from a common Lagrangian it is possible to write the canonical variables of
the MGT in terms of those of the ENM. (3) Through these substitutions we find that:
(i) the canonical algebra of the MGT can be derived from the canonical
algebra of the ENM and (ii) the Hamiltonian density of the ENM reduces to
the Hamiltonian density of the MGT. (4) Nevertheless, at this stage the Gauss functions $G_A$
in the ENM are not constraints.  In this way we need to impose additional conditions in
order to recover gauge invariance. (5) These are realized by demanding  the Gauss functions to be zero at some initial time and by recognizing that
the dynamics of the ENM is consistent with this requirement, so that they
become zero for all time. In this way, they can be added to the Hamiltonian
density of the ENM as first class constraints, thus fully recovering the gauge invariance of the MGT. As we mentioned above, since the ENM is not gauge invariant, the conservation of an external current is not guaranteed {\it a priori}, therefore it has to be imposed as an additional condition. We will show that in some cases current conservation follows from the imposition of the Gauss constraints as initial conditions.
\subsection{Solving the coordinate $\theta_1$}
\label{x1}
In this case, we solve the constraint (\ref{constriccion}) as 
\begin{equation}
\theta_{1}=f(\theta_{\bar{l}},\lambda _{A}),\quad \quad \quad \quad \bar{l}=2,...,n,
\label{constriction2}
\end{equation}
which yields
\begin{equation}
\quad \quad \quad \dot{\theta}_{1}=\;\frac{\partial f}{\partial \theta_{\bar{l}}}\dot{\theta
}_{\bar{l}}+\frac{\partial f}{\partial \lambda _{A}}\dot{\lambda}_{A},=f_{\theta_{
\bar{l}}}\dot{\theta}_{\bar{l}}+f_{\lambda _{A}}\dot{\lambda}_{A},
\label{hamiltonianconstric}
\end{equation}
where the time derivative is denoted by an overdot. Substituting the
relations (\ref{constriction2}) and (\ref{hamiltonianconstric}) directly in
the Lagrangian density (\ref{lag2}), we obtain 
\begin{equation}
\bar{\mathcal{L}}(\theta_{\bar{l}},\lambda _{A},\dot{\theta}_{\bar{l}},\dot{\lambda}
_{A})=\mathcal{L}(\theta_{1}(\theta_{\bar{l}},\lambda _{A}),\theta_{\bar{l}},\lambda _{A},
\dot{\theta}_{1}(\theta_{\bar{l}},\lambda _{A},\dot{\theta}_{\bar{l}},\dot{\lambda}_{A}),\;
\dot{\theta}_{\bar{l}}),
\end{equation}
\subsubsection{Hamiltonian and canonical algebra}
In this section we {make} explicit the relation among the canonical
coordinates of the ENM and those of the MGT. Using {such relation} we show that: (i)
the canonical algebra of the MGT is derived from the canonical algebra of
the ENM and (ii) the form of the canonical Hamiltonian density of the MGT,
given in the previous section, is obtained from the canonical Hamiltonian
density of the ENM.

In this case the independent coordinates of the ENM are $\theta_{\bar{l}}$ and $
\lambda _{A}$. We note that, the quantities $\dot{\lambda}_{A}$ appear only
as the result of imposing the constraint (\ref{constriction2}) by means of
the substitution of the velocity $\dot{\theta}_{1}$. Therefore, the momenta
associated to the $\lambda _{A}$ variables are not zero in the ENM. Note that in the ENM, 
$\theta_{1}$ and $\dot{\theta}_{1}$ are just labels to specify a particular
combination of the coordinates $\theta_{\bar{l}}$ and $\lambda _{A}$ and
velocities $\dot{\theta}_{\bar{l}}$ and $\dot{\lambda}_{A} $. After the
substitutions (\ref{constriction2}) and (\ref{hamiltonianconstric}), the Lagrange densities $\bar{
\mathcal{L}}$ and $\mathcal{L}$ have different functional form; however, let
us emphasize that the labels $\theta_{1}$ and $\dot{\theta}_{1}$ in the ENM allow us
to write, for example,
\begin{equation}
\frac{\partial \bar{\mathcal{L}}}{\partial \lambda _{A}}=\frac{\partial 
\mathcal{L}}{\partial \theta_{1}}\frac{\partial \theta_{1}}{\partial \lambda _{A}}+
\frac{\partial \mathcal{L}}{\partial \lambda _{A}},
\end{equation}
making use of the chain rule, with \ $\mathcal{L}=\mathcal{L}(\theta_{l},\lambda
_{A},\dot{\theta}_{l})$. Relations of this kind will prove very useful to compare
the Hamiltonian structure of the ENM with that of the MGT.

The canonical momenta for the ENM are given by 
\begin{eqnarray}
\bar{\Pi}_{\bar{l}}^{\theta} &\equiv &\frac{\partial \bar{\mathcal{L}}}{\partial 
\dot{\theta}_{\bar{l}}}=\left[ \frac{\partial \mathcal{L}}{\partial \dot{\theta}_{\bar{
l}}}\right] _{\theta_{1}=f}+\left[ \frac{\partial \mathcal{L}}{\partial \dot{\theta}
_{1}}\right] _{\theta_{1}=f}\frac{\partial \dot{\theta}_{1}}{\partial \dot{\theta}_{\bar{l}}
}=\left[ \frac{\partial \mathcal{L}}{\partial \dot{\theta}_{\bar{l}}}\right]
_{\theta_{1}=f}+\left[ \frac{\partial \mathcal{L}}{\partial \dot{\theta}_{1}}\right]
_{\theta_{1}=f}f_{\theta_{\bar{l}}}\,,\quad  \label{momenta2} \\
\quad \bar{\Pi}_{A}^{\lambda } &\equiv &\frac{\partial \bar{\mathcal{L}}}{
\partial \dot{\lambda}_{A}}=\left[ \frac{\partial \mathcal{L}}{\partial \dot{
\theta}_{1}}\right] _{\theta_{1}=f}\frac{\partial \dot{\theta}_{1}}{\partial \dot{\lambda}
_{A}}=\left[ \frac{\partial \mathcal{L}}{\partial \dot{\theta}_{1}}\right]
_{\theta_{1}=f}f_{\lambda _{A}},  \label{momenta3}
\end{eqnarray}
where we have obtained $\partial \dot{\theta}_{1}/\partial \dot{\theta}_{\bar{l}}$ and 
$\partial \dot{\theta}_{1}/\partial \dot{\lambda}_{A}$ from (\ref
{hamiltonianconstric}). The notation is $f_y=\partial f/\partial y$. We
assume that from Eqs. (\ref{momenta2}) we can invert the velocities $\dot{\theta}
_{\bar{l}}$ in terms of the momenta $\bar{\Pi}_{\bar{l}}^{\theta}$, so these
equations do not define constraints of the theory. On the {other hand}, after eliminating $\left[ \partial \mathcal{L}/\partial 
\dot{\theta}_{1}\right] _{\theta_{1}=f}$  in favor of $\bar{\Pi}_{1}^{\lambda }$, the Eqs.
(\ref{momenta3}) provide $(K-1)$
primary constraints which we choose as 
\begin{equation}
\phi _{\bar{A}}^{1}=\bar{\Pi}_{\bar{A}}^{\lambda }-\frac{\bar{\Pi}
_{1}^{\lambda }}{f_{\lambda _{1}}}f_{\lambda _{\bar{A}}}\approx 0,\;\;\;\;\;
\bar{A}=2,3,...,K.
\end{equation}
Let us recall that 
\begin{equation}
\frac{\partial \mathcal{L}}{\partial \dot{\theta}_{1}}\equiv \Pi
_{1}^{\theta},\;\;\;\;\;\;\frac{\partial \mathcal{L}}{\partial \dot{\theta}_{\bar{l}}}
\equiv \Pi _{\bar{l}}^{\theta},  \label{GTMOM}
\end{equation}
define the corresponding canonically conjugated momenta in the MGT {arising} from the
Lagrangian density $\mathcal{L}=\mathcal{L}(\theta_{l},\lambda _{A},\dot{\theta}_{l})$. 
In this way, when substituting $\theta_{1}=f(\theta_{\bar{l}},\lambda _{A})$ in the
expressions for $\frac{\partial \mathcal{L}}{\partial \dot{\theta}_{1}}$ and $
\frac{\partial \mathcal{L}}{\partial \dot{\theta}_{\bar{l}}}$ of Eqs. (\ref
{momenta2}), $\Pi _{1}^{\theta}\;$and $\Pi _{\bar{l}}^{\theta}\;$become just labels
used to rewrite a particular combination of coordinates and velocities in
the ENM. However, when we reinstate the notation in terms of $\theta_{1}$ and $
\dot{\theta}_{1}$, i.e. when going back to $\mathcal{L}=\mathcal{L}(\theta_{l},\lambda
_{A},\dot{\theta}_{l}),$ they recover their definition as the canonically
conjugated momenta corresponding to the MGT.

In other words, the relations (\ref{momenta2}) allow us to relate the
canonically conjugated momenta of the ENM (labeled as $\bar{\Pi}$) with
those of the MGT (labeled as $\Pi $) in the following way. 
\begin{equation}
\bar{\Pi}_{\bar{l}}^{\theta}=\Pi _{\bar{l}}^{\theta}+\Pi _{1}^{\theta}f_{\theta_{\bar{l}
}}\,,\quad \quad \quad \quad \bar{\Pi}_{A}^{\lambda }=\Pi _{1}^{\theta}f_{\lambda
_{A}}.  \label{momentaNG}
\end{equation}
In the ENM the coordinates and canonical momenta ($\theta_{\bar{l}},\lambda _{A},\bar{\Pi}_{
\bar{l}}^{\theta},\bar{\Pi}_{A}^{\lambda }$) satisfy the non-zero canonical PBs 
\begin{equation}
\{\theta_{\bar{\imath}},\bar{\Pi}_{\bar{j}}^{\theta}\}=\delta _{{\bar{\imath}}{\bar{j}}
},\quad \quad \quad \{\lambda _{A},\bar{\Pi}_{B}^{\lambda }\}=\delta _{AB}.
\label{commutation_constric}
\end{equation}
From (\ref{momentaNG}) we can express the labels $\Pi _{1}^{\theta}$ and $\Pi _{
\bar{l}}^{\theta}$ in terms of the canonical momenta $\bar{\Pi}_{\bar{l}}^{\theta}$
and $\bar{\Pi}_{l}^{\lambda }$ of \ the ENM, as 
\begin{equation}
\Pi _{\bar{l}}^{\theta}=\bar{\Pi}_{\bar{l}}^{\theta}-\bar{\Pi}_{1}^{\lambda }\frac{
f_{\theta_{\bar{l}}}}{f_{\lambda _{1}}}\,,\quad \quad \quad \quad \Pi _{1}^{\theta}=
\frac{\bar{\Pi}_{1}^{\lambda }}{f_{\lambda _{1}}}.  \label{momentaTTNM}
\end{equation}
In the \ref{Apendice1}, we show that when considering $\Pi
_{1}^{\theta}$ and $\Pi _{\bar{l}}^{\theta}$ as labels in terms of the canonical
variables of the ENM $(\theta_{\bar{l}},\bar{\Pi}_{\bar{l}}^{\theta},\lambda _{A},\bar{
\Pi}_{A}^{\lambda })$, we obtain the following PB algebra 
\begin{equation}
\{\theta_{i},\theta_{j}\}=0,\quad \quad \quad \{\Pi _{i}^{\theta},\Pi _{j}^{\theta}\}=0,\quad
\quad \quad \{\theta_{i},\Pi _{j}^{\theta}\}=\delta _{ij},
\end{equation}
by using the canonical algebra (\ref{commutation_constric}) of the ENM.

The calculation of the Hamiltonian density $\mathcal{H}_{c}^{ENM}$ of the
ENM gives 
\begin{eqnarray}
\mathcal{H}_{c}^{ENM} &=&\dot{\lambda}_{A}\bar{\Pi}_{A}^{\lambda }+\dot{\theta}_{
\bar{l}}\bar{\Pi}_{\bar{l}}^{\theta}-\bar{\mathcal{L}},  \notag \\
&=&\dot{\lambda}_{A}\Pi _{1}^{\theta}f_{\lambda _{A}}+\dot{\theta}_{\bar{l}}(\Pi _{
\bar{l}}^{\theta}+\Pi _{1}^{\theta}f_{\theta_{\bar{l}}})-\bar{\mathcal{L}},  \notag \\
&=&\Pi _{1}^{\theta}(f_{\lambda _{A}}\dot{\lambda}_{A})+\dot{\theta}_{\bar{l}}(\Pi _{
\bar{l}}^{\theta}+\Pi _{1}^{\theta}f_{\theta_{\bar{l}}})-\bar{\mathcal{L}},  \notag \\
&=&\Pi _{1}^{\theta}(\dot{\theta}_{1}-f_{\theta_{\bar{l}}}\dot{\theta}_{\bar{l}})+\dot{\theta}_{\bar{l
}}(\Pi _{\bar{l}}^{\theta}+\Pi _{1}^{\theta}f_{\theta_{\bar{l}}})-\bar{\mathcal{L}},  \notag
\\
&=&\dot{\theta}_{1}\Pi _{1}^{\theta}+\dot{\theta}_{\bar{l}}\Pi _{\bar{l}}^{\theta}-\bar{\mathcal{
L}},  \notag \\
&=&\dot{\theta}_{k}\Pi _{k}^{\theta}-\bar{\mathcal{L}},\quad \quad \quad \quad \quad
\quad \quad \quad \quad \quad (k=1,...,n),  \label{HCENM}
\end{eqnarray}
where we have substituted $f_{\lambda _{A}}\dot{\lambda}_{A}$ from Eq. (\ref
{hamiltonianconstric}) and $(\bar{\Pi}_{\bar{l}}^{\theta},\bar{\Pi}_{A}^{\lambda
})$ from Eq.\ (\ref{momentaNG}).

Undoing the substitution (\ref{constriction2}), that is to say, inserting
back the original variables $\theta_{1}$ and $\dot{\theta}_{1}$ in Eq. (\ref{HCENM}),
we realize that $\bar{\mathcal{L}}$ reduces to $\mathcal{L}$ and that $\Pi
_{k}^{\theta}$ are the corresponding canonically conjugated momenta of the MGT,
according to Eqs. (\ref{GTMOM}). In this way, the Hamiltonian density (\ref
{HCENM}) has the same form that {${\mathcal H}_{c}$} given in Eq. (\ref
{hamiltonianocanonic}) for the MGT.

Next, we consider the extended Hamiltonian density which is given by 
\begin{eqnarray}
\mathcal{H}_{E}^{ENM} &=&\mathcal{H}_{c}^{ENM}+\mu _{\bar{A}}\phi _{\bar{A}
}^{1},  \notag \\
&=&\mathcal{H}_{F}+\lambda _{A}G_{A}+\theta_lJ_l+\mu _{\bar{A}}\left( \bar{\Pi}_{\bar{A}
}^{\lambda }-\frac{\bar{\Pi}_{1}^{\lambda }}{f_{\lambda _{1}}}f_{\lambda _{
\bar{A}}}\right) ,
\end{eqnarray}
where we have used that $\mathcal{H}_{c}^{ENM}$ has the same form that $
\mathcal{H}_{c}$ in (\ref{hamiltonianocanonic}) and $\mu _{\bar{A}}$ are
arbitrary functions. In the \ref{Apendice1}, we show that $\{\phi _{
\bar{A}}^{1},\theta_{l}\}=0$, $\{\phi _{\bar{A}}^{1},\Pi _{l}^{\theta}\}=0$ and $
\{\phi _{\bar{A}}^{1},\phi _{\bar{B}}^{1}\}=0$. Using the previous results,
the time evolution condition of the primary constraints leads to 
\begin{equation}
\dot{\phi}_{\bar{A}}^{1}=\{\phi _{\bar{A}}^{1},{H}
_{E}^{ENM}\}={\int d^3y \{\phi _{\bar{A}}^{1},\lambda _{\bar{B}}(y)\}G_{\bar{B}}(y)}=\left( G_{
\bar{A}}-G_{1}\frac{f_{\lambda _{\bar{A}}}}{f_{\lambda _{1}}}\right) ,
\end{equation}
where we identify the secondary constraints 
\begin{equation}
\phi _{\bar{A}}^{2}=f_{\lambda _{\bar{A}}}-\frac{G_{\bar{A}}}{G_{1}}
f_{\lambda _{1}}\approx 0.  \label{constrsecun}
\end{equation}
Next, we calculate the time evolution of $\phi _{\bar{A}}^{2}$, with the
result 
\begin{eqnarray}
\dot{\phi}_{\bar{A}}^{2}=\{\phi _{\bar{A}}^{2},{{H}}_{E}^{ENM}\}
=U_{\bar{A}}-\mu _{\bar{B}}T_{\bar{B}\bar{A}}\approx 0, 
\end{eqnarray}
with $U_{\bar{A}}=\{\phi _{\bar{A}}^{2},{H}_{c}^{ENM}\}$ and $T=
\left[ T_{\bar{A}\bar{B}}\right] =\left[ \{\phi _{\bar{A}}^{1},\phi _{\bar{B}
}^{2}\}\right] $. Unless confusion arises, here and in the following we make use of the convention introduced  after Eq. (\ref{inversa}), whereby
$\mu _{\bar{B}}T_{\bar{B}\bar{A}}$ is given by $\int d^3 y \, \mu_{\bar{B}}(y)T_{\bar{B}\bar{A}}(y,x)$, for example.

In the \ref{Apendice1}, we show that it is possible to choose
particular functions $f(\theta_{\bar{l}},\lambda _{A})$ such that, the matrix $T_{
\bar{A}\bar{B}}$ is invertible, which means that $\phi _{\bar{A}}^{1}$ and $
\phi _{\bar{B}}^{2}$ are second class constraints. The arbitrary functions $
\mu _{\bar{B}}$ are fixed as 
\begin{equation}
\mu _{\bar{A}}=(T^{-1})_{\bar{B}\bar{A}}U_{\bar{B}}.
\end{equation}
The Dirac method stops and the ENM only has the following $2(K-1)\;$second
class constraints 
\begin{equation}
\phi _{\bar{A}}^{1}=\bar{\Pi}_{\bar{A}}^{\lambda }-\frac{\bar{\Pi}
_{1}^{\lambda }}{f_{\lambda _{1}}}f_{\lambda _{\bar{A}}},\quad \quad \quad
\phi _{\bar{A}}^{2}=f_{\lambda _{\bar{A}}}-f_{\lambda _{1}}\frac{G_{\bar{A}}
}{G_{1}}.  \label{constspacelike}
\end{equation}
Thus, the number of DOF of the ENM is 
\begin{equation}
\#\text{DOF}=\frac{1}{2}[2(n-1+K)-2(K-1)]=n.
\label{dofx1a}
\end{equation}
As previously emphasized, the ENM is not a gauge invariant theory, it has
only second class constraints, and the number of degrees of freedom is not
the same as in the MGT. From Eq. (\ref{dofgauge}), we can observe that the
ENM has $K$ degrees of freedom more than the\ MGT, so that, if we want to
establish an equivalence, we will have to impose $K$ additional constraints
to the ENM.

The next step is to set strongly equal zero the second class constraints (\ref{constspacelike}). 
To this end we introduce the corresponding DBs and
further calculate them among the remaining variables. We require the matrix
constructed with the PBs of the constraints 
\begin{equation}
M=\left[ 
\begin{array}{cc}
\left[ R_{\bar{A}\bar{B}}\right] & \left[ T_{\bar{A}\bar{B}}\right] \\ 
-\left[ T_{\bar{B}\bar{A}}\right] & \;\left[ S_{\bar{A}\bar{B}}\right]
\end{array}
\right] =\left[ 
\begin{array}{cc}
R & T \\ 
-T^{T} & \;S
\end{array}
\right] ,  \label{SL_PP_MATRIX}
\end{equation}
where 
\begin{equation}
R_{\bar{A}\bar{B}}=\left\{ \phi _{\bar{A}}^{1},\;\;\phi _{\bar{B}
}^{1}\right\} ,\;\;\;T_{\bar{A}\bar{B}}=\left\{ \phi _{\bar{A}}^{1},\;\;\phi
_{\bar{B}}^{2}\right\} ,\;\;\;\;S_{\bar{A}\bar{B}}=\left\{ \phi _{\bar{A}
}^{2},\;\;\phi _{\bar{B}}^{2}\right\} \;.  \label{SL_PP_ENTRIES}
\end{equation}
In the \ref{Apendice1} we show that $R_{\bar{A}\bar{B}}=0$. The
inverse\ matrix becomes 
\begin{equation}
\;\;M^{-1}=\left[ 
\begin{array}{cc}
T^{-1}ST^{-1} & -T^{-1} \\ 
T^{-1} & 0
\end{array}
\right] .
\end{equation}
The DBs are 
\begin{eqnarray}
\{\mathcal{A},\mathcal{B}\}_{D} &=&\{\mathcal{A},\mathcal{B}\}-\{\mathcal{A}
,\phi _{\bar{A}}^{1}\}(T^{-1}ST^{-1})_{\bar{A}\bar{B}}\{\phi _{\bar{B}}^{1},
\mathcal{B}\}  \notag  \label{bracketDspace1} \\
&&+\{\mathcal{A},\phi _{\bar{A}}^{1}\}(T^{-1})_{\bar{A}\bar{B}}\{\phi _{\bar{
B}}^{2},\mathcal{B}\}-\{\mathcal{A},\phi _{\bar{A}}^{2}\}(T^{-1})_{\bar{A}
\bar{B}}\{\phi _{\bar{B}}^{1},\mathcal{B}\},
\end{eqnarray}
which again leads to the result 
\begin{equation}
\{\mathcal{A}(x),\mathcal{B}(y)\}_{D}=\{\mathcal{A}(x),\mathcal{B}(y)\},
\label{DBEPBSL}
\end{equation}
for the coordinates $\theta_{j}$ and momenta $\Pi _{j}^{\theta}\;$written in terms of
the canonical variables $\theta_{\bar{l}},\;\bar{\Pi}_{\bar{l}}^{\theta}\;$of the ENM.
The above result arises from the fact that each of the additional PBs in (\ref
{bracketDspace1}) includes a contribution containing $\phi _{\bar{A}}^{1}$,
which has zero PB with $\theta_{j}$ and $\Pi _{j}^{\theta}$, according to the results in \ref{Apendice1}.

{Using} the transformations {given in } (\ref{constriction2}) and 
(\ref{momentaTTNM}) we can rewrite all quantities of the ENM in terms of the
labels $\theta_{j}$ and $\Pi _{j}^{\theta}$. {Moreover, since}  we know the canonical algebra
that {the latter} satisfy, at some stage it becomes more convenient to employ such
labels to describe the ENM, instead of its own canonical variables. The
above analysis shows that we recover the canonical algebra (\ref{algebrafinal}) of the
MGT. Once the Lagrange multipliers $\mu _{\bar{A}}$ have been fixed and the
constraints $\phi _{\bar{A}}^{1}\;$and $\phi _{\bar{A}}^{2}\;$have been
imposed strongly, the Hamiltonian {density} of the ENM is given by 
\begin{equation}
\mathcal{H}_{E}^{ENM}=\mathcal{H}_{F}+\lambda _{A}G_{A}+\theta_lJ_l.
\label{hamiltonianofinal2}
\end{equation}

\subsubsection{Conservation of the quantities $G_{A}$}

Up to this stage we have only verified that, with the appropriate change of
variables (\ref{constriction2}) and (\ref{momentaTTNM}), the canonical
algebra of the ENM induces the canonical algebra of the related \ MGT. Under
the same transformations, the canonical Hamiltonian of the ENM adopts the
same form as that of the\ MGT. Now we explore the conditions under which the
full\ MGT emerges from the ENM. At this point, two fundamental differences
arise: (a) the quantities $\lambda _{A}$ in the extended Hamiltonian density
(\ref{hamiltonianofinal2}) are  functions of the coordinates and momenta of the MGT, i. e. 
$\lambda_A=\lambda_A(\theta_l, \Pi^\theta_k)$ instead of been 
arbitrary functions, as it is required in the Hamiltonian density of the MGT, where they should
correspond to the $\Theta _{A}$'s appearing in Eq. (\ref
{hamiltonianofinal}).  Such relations arise after imposing strongly the second class constraints (\ref{constspacelike}) of the ENM. (b) The Gauss functions $G_{A}$ in Eq. (\ref
{hamiltonianofinal2})\ are not constraints in the ENM, while in the MGT 
they should be realized as first class constraints, being the
generators of the gauge symmetry.

To deal with these issues, we study the time evolution of the quantities $
G_{A}$ under the dynamics of the ENM 
\begin{eqnarray}
\dot{G}_{A}&=&\{G_{A},{H}_{E}^{ENM}\}={{\dot J}_A}+\{G_{A},{H}_{F}\}+\int d^3y \lambda _{B}(y)\{G_{A},G_{B}(y)\} \notag \\
&&+\int d^3y ( \{G_A,\theta_l(y)\}J_l(y)+  \{G_{A},\lambda_{B}(y)\} G_{D}(y)).   \label{GMDOT}
\end{eqnarray}
Using the the equivalence between the canonical algebras, which we have already
proved, we can employ the results arising from the MGT to evaluate the
first two brackets  in Eq. (\ref{GMDOT}). From Eq. (\ref{condition3}) we
obtain 
\begin{eqnarray}
\dot{G}_{A}=\dot{J}_A+{C}_{AB}G^0_{B}&+& C_{ABC}\lambda_B G_C^0  + \int d^3y (\{G_A,\theta_l(y)\}J_l(y)+\{G_{A},\lambda
_{B}(y)\}G_{B}(y)), \nonumber \\
\dot{G}_{A} &=& {C}_{AB}G_{B}+ C_{ABC}\lambda_B G_C + \int d^3y \{G_{A},\lambda
_{B}(y)\}G_{B}(y)\nonumber \\
&+& \dot{J}_A-{C}_{AB}J_{B}- C_{ABC}\lambda_B J_C  + \int d^3y \{G^0_{A},\theta_l(y)\}J_l(y), \nonumber \\
\dot{G}_{A} &=& {C}_{AB}G_{B}+ C_{ABC}\lambda_B G_C +  \int d^3y\{G_{A},\lambda
_{B}(y)\}G_{B}(y)+(DJ)_A, \notag \\
\label{DOTGAENM1}
\end{eqnarray}
where we have used the relation $G_A^0=G_A-J_A$. Let us recall that $(DJ)_A$ was defined in Eq. (\ref{DEFDJA}). 

At this stage 
$\dot{G}_{A} \neq 0$ for two reasons: (1) the Gauss functions $G_A$ are not constraints and (2) the generalized current conservation 
$(DJ)_A=0$ is not valid due to the lack of gauge invariance in the ENM. Then, in order to have $\dot{G}_{A}=0$ it is enough  to impose the following conditions upon the ENM: (i) $(DJ)_A=0$  for all times  and (ii) $G_{A}=0$ as an initial condition, at $t=0$ for example, which also leads to ${\dot G}_A=0$ at $t=0$. In this way,  Eq. (\ref{DOTGAENM1}) yields $G_{A}=0$ for all time. 

Under the above conditions,  we can recover the MGT by adding the quantities $G_{A}$
as Hamiltonian first class constraints, $G_{A}\approx 0$,  with arbitrary
functions $N_{A}$. This requires to add $N_{A}G_{A}$ to the
Hamiltonian (\ref{hamiltonianofinal2}) and to redefine $\lambda
_{A}+N_{A}=\Theta _{A}$, which leads to 
\begin{equation}
\mathcal{H}_{E}=\mathcal{H}_{F}+\Theta _{A}G_{A}+\theta_lJ_l,
\label{HDENM1}
\end{equation}
where $\Theta _{A}$ are now arbitrary functions.

In other words, we have regained the Hamiltonian (\ref{hamiltonianofinal}) 
together with the canonical algebra (\ref{algebrafinal}) of the\ MGT.
Summarizing, the equivalence between the MGT and the ENM model can be
established only after imposing the generalized current conservation $(DJ)_A=0$ for all times, together with $G_{A}=0$ as initial
conditions in the dynamics of the ENM.

As we have previously shown, the number of DOF of the ENM is $n$, but when
the $K$ relations $G_{A}=0$ are imposed as first class constraints into the
Hamiltonian density, the remaining theory has only $n-K$ DOF, which is the same
number of DOF of the\ MGT.

\subsection{Solving the coordinate $\protect\lambda _{1}$}

A similar analysis can be performed for this case. We solve the constraint (\ref{constriccion}) as 
\begin{equation}
\lambda _{1}=g(\theta_{i},\lambda _{\bar{B}})\quad \quad \quad \bar{B}=2,3,...,K.
\label{const2}
\end{equation}
in such a way that the canonical variables for the ENM are now\ $\theta_{i},\;
\bar{\Pi}_{i}^{\theta},\;\lambda _{\bar{B}},\;\bar{\Pi}_{\bar{B}}^{\lambda }.\;$
The velocities $\dot{\lambda}_{A}$ do not appear in the Lagrangian (\ref
{Lagrangian}), and since the constraint (\ref{const2}) does not introduce
additional velocity dependent terms, we have 
\begin{equation}
\frac{\partial \bar{\mathcal{L}}}{\partial \dot{\theta}_{i}}=\bar{\Pi}
_{i}^{\theta}=\Pi _{i}^{\theta},\quad \quad \quad \frac{\partial \bar{\mathcal{L}}}{
\partial \dot{\lambda}_{\bar{B}}}=\Pi _{\bar{B
}}^{\lambda }=0,
\label{constr123a}
\end{equation}
where $\Pi _{i}^{\theta}$ and $\Pi _{\bar{B}}^{\lambda }$ \ are given by (\ref
{moments}) satisfying the same PBs indicated in (\ref{commutationfree}).

\subsubsection{Hamiltonian and canonical algebra of the ENM}

In this case, $\lambda _{1}$ and $\Pi _{1}^{\lambda }$ are not independent canonical
variables of the ENM, in fact $\lambda_1$ is a function of the remaining variables 
$\lambda _{1}=g(\theta_{i},\lambda _{\bar{B}})$
and $\Pi _{1}^{\lambda }$ is not included in the ENM. The calculation of the canonical Hamiltonian
density proceeds just as in the previous Section \ref{gaugetheory} and we
find 
\begin{eqnarray}
\mathcal{H}_{c}^{ENM} &=&\mathcal{H}_{F}+\lambda _{1}G_{1}+\lambda _{\bar{B}
}G_{\bar{B}}+\theta_lJ_l,  \notag \\
&=&\mathcal{H}_{F}\;+g(\theta_{m},\lambda _{\bar{B}})G_{1}+\lambda _{\bar{B}}G_{
\bar{B}}+\theta_lJ_l.
\end{eqnarray}
The theory contains $(K-1)$ primary constraints $\phi _{\bar{B}}^{1}=\bar{\Pi
}_{\bar{B}}^{\lambda }\approx 0$, $\bar{B}=2,3,...,K$. The extended
Hamiltonian is given by 
\begin{eqnarray}
\mathcal{H}_{E}^{ENM} &=&\mathcal{H}_{c}^{ENM}+\mu _{\bar{B}}\bar{\Pi}_{\bar{
B}}^{\lambda }, \\
&=&\mathcal{H}_{F}+g(\theta_{m},\lambda _{\bar{B}})G_{1}+\lambda _{\bar{B}}G_{
\bar{B}}+\theta_lJ_l+\mu _{\bar{B}}\bar{\Pi}_{\bar{B}}^{\lambda },  \notag
\end{eqnarray}
where $\mu _{\bar{B}}$ are arbitrary functions. The time
evolution condition for the primary constraints gives 
\begin{eqnarray}
\dot{\phi}_{\bar{B}}^{1} =\{\phi _{\bar{B}}^{1},{H}_{E}^{ENM}\}
=\{\Pi _{\bar{B}}^{\lambda },{H}_{E}^{ENM}\}
=-(g_{\lambda _{\bar{B}}}G_{1}+G_{\bar{B}}), 
\end{eqnarray}
yielding $(K-1)$ secondary constraints, which we write as 
\begin{equation}
\phi _{\bar{B}}^{2}=g_{\lambda _{\bar{B}}}+\frac{G_{\bar{B}}}{G_{1}}\approx
0.
\end{equation}
From the time evolution condition for the secondary constraints, we obtain 
\begin{eqnarray}
\dot{\phi}_{\bar{B}}^{2}&=&\{\phi _{\bar{B}}^{2},{H} 
_{E}^{ENM}\}=\{\phi _{\bar{B}}^{2},{H}_{c}^{ENM}\} + \nonumber \\
&& + {\int d^3y \,\mu _{\bar{A}}(y)\{\phi _{\bar{B
}}^{2},\phi _{\bar{A}}^{1}(y)\}}=W_{\bar{B}}-\mu _{\bar{A}}X_{\bar{B}\bar{A}}\approx 0,
\end{eqnarray}
where $W_{\bar{B}}=\{\phi _{\bar{B}}^{2},\mathcal{H}_{c}^{ENM}\}$ and $X_{
\bar{A}\bar{B}}=\{\phi _{\bar{A}}^{1},\phi _{\bar{B}}^{2}\}$. Once again, in the \ref{Apendice2} we show that there exist particular
functions $g(\theta_{m},\lambda _{\bar{B}})$ such that the matrix $X_{\bar{A}\bar{
B}}$ is invertible, therefore, the Lagrange multipliers $\mu _{\bar{B}}$ are
fixed as 
\begin{equation}
\mu _{\bar{B}}=(X^{-1})_{\bar{B}\bar{A}}W_{\bar{A}}.
\end{equation}
The existence of $(X^{-1})_{\bar{B}\bar{A}}$ guarantees that $\phi _{\bar{B}
}^{1}$ and $\phi _{\bar{B}}^{2}$ are second class constraints and that the
Dirac method stops. The number of DOF is 
\begin{equation}
\#\text{DOF}=\frac{1}{2}[2(n+K-1)-2(K-1)]=n.
\label{dof1ab}
\end{equation}
As usual, we set strongly equal zero the constraints $\phi _{\bar{B}}^{1}$
and $\phi _{\bar{B}}^{2}$ to subsequently introduce the corresponding
DBs. We require the matrix constructed with the PBs of the constraints 
\begin{equation}
M=\left[ 
\begin{array}{cc}
\left[ R_{\bar{A}\bar{B}}\right] & \left[ X_{\bar{A}\bar{B}}\right] \\ 
-\left[ X_{\bar{B}\bar{A}}\right] & \;\left[ S_{\bar{A}\bar{B}}\right]
\end{array}
\right] =\left[ 
\begin{array}{cc}
R & X \\ 
-X^{T} & \;S
\end{array}
\right] , 
\end{equation}
where 
\begin{equation}
R_{\bar{A}\bar{B}}=\left\{ \phi _{\bar{A}}^{1},\;\;\phi _{\bar{B}
}^{1}\right\} ,\;\;\;X_{\bar{A}\bar{B}}=\left\{ \phi _{\bar{A}}^{1},\;\;\phi
_{\bar{B}}^{2}\right\} ,\;\;\;\;S_{\bar{A}\bar{B}}=\left\{ \phi _{\bar{A}
}^{2},\;\;\phi _{\bar{B}}^{2}\right\} \,.\; 
\end{equation}
Again, in the \ref{Apendice2} we show that $R_{\bar{A}\bar{B}
}=0$. The inverse matrix $M^{-1}$ is given by 
\begin{equation}
\;\;M^{-1}=\left[ 
\begin{array}{cc}
X^{-1}SX^{-1} & -X^{-1} \\ 
X^{-1} & 0
\end{array}
\right] .
\end{equation}
The DBs are 
\begin{eqnarray}
\{\mathcal{A},\mathcal{B}\}_{D} &=&\{\mathcal{A},\mathcal{B}\}-\{\mathcal{A}
,\phi _{\bar{A}}^{1}\}(X^{-1}SX^{-1})_{\bar{A}\bar{B}}\{\phi _{\bar{B}}^{1},
\mathcal{B}\}  \notag  \label{bracketDspace} \\
&&+\{\mathcal{A},\phi _{\bar{A}}^{1}\}(X^{-1})_{\bar{A}\bar{B}}\{\phi _{\bar{
B}}^{2},\mathcal{B}\}-\{\mathcal{A},\phi _{\bar{A}}^{2}\}(X^{-1})_{\bar{A}
\bar{B}}\{\phi _{\bar{B}}^{1},\mathcal{B}\},
\end{eqnarray}
which again leads to the result 
\begin{equation}
\{\mathcal{A}(\mathbf{x}),\mathcal{B}(\mathbf{y})\}_{D}=\{\mathcal{A}(
\mathbf{x}),\mathcal{B}(\mathbf{y})\}, 
\end{equation}
for the final variables $\theta_{j}$ and $\Pi _{j}^{\theta}$. The above result arises
from the fact that each one of the additional PBs in 
(\ref{bracketDspace}) include a contribution with $\phi _{\bar{A}}^{1}$, 
which has zero PB with $\theta_{j}$ and $\Pi _{j}^{\theta}$, according to the \ref{Apendice2}. In other \ words, we recover the canonical
algebra (\ref{algebrafinal}), together the final extended Hamiltonian
density 
\begin{equation}
\mathcal{H}_{E}^{ENM}=\mathcal{H}_{F}+\lambda _{A}G_{A}+\theta_lJ_l.
\label{hamiltonianofinal3}
\end{equation}

\subsubsection{Conservation of the quantities $G_{A}$}

As in the case of Subsection \ref{x1}, we have to deal with the issue that neither the $
G_{A}$'s are constraints, nor the $\lambda _{A}$'s are arbitrary functions
in the extended Hamiltonian density (\ref{hamiltonianofinal3}) of the ENM.
However, {the transformation of the Gauss functions $G_{A}$ into constraints}  proceeds in the same way as in  the
previous case. In fact, { after employing the equivalence between the canonical algebras of
the MGT and the ENM in the calculation},  the dynamics of the ENM yields { again Eq.(\ref{DOTGAENM1})}.
Once more, in order to recover the  MGT, it is enough to
impose the generalized current conservation $(DJ)_A=0$ for all times and $G_{A}=0$ as initial conditions, since the dynamics of the
ENM guarantees that $G_{A}(t)=0$ for all time.  Therefore, we can include the quantities $\ G_{A}$ as
constraints in the extended Hamiltonian (\ref{hamiltonianofinal3}), through
arbitrary functions $N_{A}$, 
by adding $N_{A}G_{A}$ and redefining $\lambda _{A}+N_{A}=\Theta
_{A}$. We obtain 
\begin{equation}
\mathcal{H}_{E}=\mathcal{H}_{F}+\Theta _{A}G_{A}+\theta_lJ_l,
\label{HDENM2}
\end{equation}
where now $\Theta _{A}\;$are arbitrary functions. In this way we recover the
extended Hamiltonian density (\ref{hamiltonianofinal}) together the
canonical algebra (\ref{algebrafinal}) of the MGT. The conditions for the
equivalence between the\ MGT$\;$and the ENM have been established, yielding
the same results as in the previous case of Subsection \ref{x1}.

We close this Subsection with a comment related to the r\^ole of the gauge fixing in the MGT  when achieving its equivalence with the ENM. Let us recall that such equivalence was obtained basically in two steps:
(i) we proved that the canonical algebra of the ENM yields the canonical algebra Eq. (\ref{algebrafinal}) of the MGT and (ii) after the imposition of suitable conditions (Gauss laws and current conservation) in the ENM we were able to show that the corresponding  Hamiltonian densities (\ref{HDENM1}) and  (\ref{HDENM2}) have exactly the same form as the  Hamiltonian density (\ref{algebrafinal}) of the MGT. The only explicit gauge fixing in the MGT was to set strongly the first class constraints  $\Phi^1_A=\Pi^\lambda_A \approx 0$ by adding the condition $\Phi^3_A=\lambda_A-\Theta_A$, in order to eliminate the variables $\lambda_A$ and $\Pi^\lambda_A$. The canonical algebra (\ref{algebrafinal}), together with the Hamiltonian density (\ref{hamiltonianofinal}) of the MGT were obtained at this stage. Anyway, the above gauge fixing is completely general because the arbitrary functions $\Theta_A$ remained unspecified and can only be determined once a further gauge fixing of the remaining first class constraints $G_A$ is performed. This has not been done and both the canonical algebra and the Hamiltonian densities of the MGT and the ENM coincide while keeping $G_A$ as first class constraints. In other words, the equivalence has been proved for an arbitrary gauge fixing in the MGT, after the variables $\lambda_A$ and $\Pi^\lambda_A$ were decoupled. Under the general conditions stated in the \ref{Apendice1} and the \ref{Apendice2}
such equivalence is completely independent of  the specific constraint $F(\theta_l, \lambda_A)=0$ which defines the ENM.

\section{Examples}

\label{examples} In this section, we present some examples where the above
ideas can be applied. In each case we outline a brief description of the
corresponding MGT, showing that the conditions (\ref{hamiltonianocanonic}), 
(\ref{condition2}) and (\ref{condition3}) in Section \ref{gaugetheory}
hold. These requirements are enough to establish an equivalence with an
arbitrary ENM, which would be defined by the same Lagrangian of the\ MGT$\;$
plus a constraint $F(\theta_{l},\lambda _{A})=0$. This constraint leads
to a new theory which is different of the MGT. However, after imposing
suitable conditions we can recover the original MGT as explained in
the Section \ref{NGmodel}.

\subsection{A Mechanical Model}

As a first example, we consider  a mechanical model \cite{Dayi} defined by the
Lagrangian 
\begin{equation}
\mathcal{L}=\frac{1}{2}(\dot{x}^{2}+\dot{y}^{2})-\lambda (x\dot{y}-y\dot{x})+
\frac{1}{2}\lambda ^{2}(x^{2}+y^{2})-V(x^{2}+y^{2}),
\label{MECMOD}
\end{equation}
which is invariant under the following gauge transformations, with parameter $\epsilon(t)$,
\begin{equation}
\delta\, x=\epsilon(t) \,y, \qquad \delta\, y=-\epsilon(t)\,x, \qquad  
\delta \, \lambda = {\dot \epsilon}(t), 
\end{equation}
and describes time dependent rotations around the $z$ axis.
It can be easily proved that the canonical Hamiltonian is given by 
\begin{eqnarray}
\mathcal{H}_{c} &=&\frac{\Pi _{x}^{2}}{2}+\frac{\Pi _{y}^{2}}{2}
+V(x^{2}+y^{2})+\lambda (x\Pi _{y}-y\Pi _{x}), \\
&=&\mathcal{H}_{F}(x,y,\Pi _{x},\Pi _{y})+\lambda (x\Pi _{y}-y\Pi _{x}), 
\notag
\end{eqnarray}
where 
\begin{equation}
\mathcal{H}_{F}=\frac{\Pi _{x}^{2}}{2}+\frac{\Pi _{y}^{2}}{2}
+V(x^{2}+y^{2}),\quad \quad \quad \Pi _{a}=\frac{\partial \mathcal{L}}{
\partial \dot{a}},\quad \quad \quad a=x,y.
\end{equation}
The model has only the primary constraint $\Phi _{1}=\Pi _{\lambda }=0$,
which leads to the secondary constraint $\Phi _{2}=x\Pi _{y}-y\Pi _{x}=0$.
We note that, $\Phi _{2}=\Phi _{2}(x,y,\Pi _{x},\Pi _{y})$, i.e. it is $
\lambda $ independent. Following Section \ref{gaugetheory} we identify $
\lambda \rightarrow \lambda $ and $G_{A\;\;}\rightarrow \Phi _{2}=x\Pi _{y}-y\Pi _{x}$. The
conditions $\{\Phi _{2},\mathcal{H}_{F}\}=0$ and $\{\Phi _{2},\Phi _{2}\}=0$
hold, therefore, there are no {additional} constraints in the theory. The conditions
(\ref{hamiltonianocanonic}), (\ref{condition2}) and (\ref{condition3})
demanded in Section \ref{gaugetheory} are fulfilled.

The explicit equivalence between the MGT and the corresponding ENM arising from this mechanical model will be proved in the \ref{Mechanics1a}. This is a detailed example which shows how does our procedure works.

\subsection{Yang-Mills Theory}

This case was studied in Ref. \citen{NANM_ES_UR}, where the non abelian Nambu's
model was defined by the constraint $A_{\mu}^a A^{a\mu}=n^2M^2$, where $\mu$
and $a$ are indices in the Lorentz and group spaces, respectively, $n^2=n_\beta n^\beta=0,\pm1$
and $M^2$ is a constant. The space-like case presented in \cite{NANM_ES_UR}  is
a particular example what we have proved in this work. A brief review of the
Yang-Mills case starts with the standard Lagrangian density 
\begin{equation}
\mathcal{L}(A_\mu^a)=-\frac{1}{4}F_{\mu\nu}^aF^{a\mu\nu}-A_\mu^aJ^{a\mu},
\label{LYM}
\end{equation}
which produces the Hamiltonian density
\begin{eqnarray}
\mathcal{H}_{c} &=&\frac{1}{2}(\mathbb{E}^{2}+\mathbb{B}^{2})-A_{0}^{a\;\;}(
\partial _{i}E_{i}-J_{0})^{a\;\;}+A_{i}^{a}J^{ai},\notag \\
&=&\mathcal{H}_{F}(A_{i},E_{i})-A_{0}^{a\;\;}(\partial
_{i}E_{i}-J_{0})^{a} +A_{i}^{a}J^{ai}, 
\end{eqnarray}
where 
\begin{equation}
\mathcal{H}_{F}=\frac{1}{2}(\mathbb{E}^{2}+\mathbb{B}^{2}).
\end{equation}
The non trivial gauge transformations in the Lagrangian (\ref{LYM}) are $\delta A_\mu^a=D_\mu\Lambda^a(x)$, where $D_\mu$ is the covariant derivative. The primary constraints are $\Pi _{0}^{a\;\;}=0$, which lead to the
secondary constraints $G_{a}=(\partial
_{i}E_{i}-J_{0})_{a\;\;}=G_{a}(A_{i}^{b},E_{i}^{b})$. Following 
Section \ref{gaugetheory}, we identify $A_{0}^{a\;\;}\rightarrow \lambda
_{A\;\;}$ and $G_{a\;\;}\rightarrow G_{A}$. It is well known that $\{G_{a},
\mathcal{H}_{F}\}=0$ and $\{G_{a},G_{b}\}\approx 0$, therefore, there are no
more constraints. The conditions (\ref{hamiltonianocanonic}), (\ref
{condition2}) and (\ref{condition3}) are satisfied.

As we previously mentioned, in some cases current conservation follows from 
the imposition of Gauss constraints as initial conditions, as it happens in the non-Abelian Nambu 
model, for example. We clarify this issue in the case discussed in Ref. \citen{NANM_ES_UR}, defined by
\begin{equation}
\mathcal{L}(A_\mu)=-\frac{1}{4}F_{\mu\nu}^aF^{a\mu\nu}-A_\mu^aJ^{a\mu}, \quad\quad\quad A_{\mu}^a A^{a\mu}=n^2M^2.
\end{equation}
The equations of motion for the space-like case are given by
\begin{equation}
\mathcal{E}^{ia}-\mathcal{E}^{31}\frac{A_i^a}{A_3^1}=0, \quad\quad i \neq 3, \quad a\neq 1.
\label{NANMeq1}
\end{equation}
\begin{equation}
\mathcal{E}^{0a}-\mathcal{E}^{31}\frac{A_0^a}{A_3^1}=0,
\label{NANMeq2}
\end{equation}

with the notation 

\begin{equation}
\mathcal{E}^{\nu a}=(D_\mu F^{\mu\nu}-J^\nu)^a.
\end{equation}

From the dynamics of the non-Abelian Nambu model the time evolution of the Gauss functions $\Omega^a=\mathcal{E}^{0a}=(D_i E_i- J_0)^a$ is

\begin{equation}
\dot{\Omega}^a=-g C^{abc}A_0^b \Omega^c - D_\mu J^{\mu a} + D_k \bigg(\bigg(\frac{A_k^a}{A_0^1}\bigg)\Omega^1\bigg),
\label{NANMGauss}
\end{equation}

Imposing the Gauss constraints as initial conditions ($\Omega^a(t=t_0)=0$) upon Eqs. (\ref{NANMeq1})-(\ref{NANMeq2}), the standard Yang-Mills equations of motion ($\mathcal{E}^{\nu a}=0$) are recovered and they are valid at $t=t_0$. As a consequence of the antisymmetry of Maxwell tensor, the relation 
\begin{equation}
0=D_\nu \mathcal{E}^{\nu a}= D_\nu (D_\mu F^{\mu\nu}-J^\nu)^a= -D_\nu J^{\nu a}=0, \quad \quad (\textrm{at } t=t_0),
\end{equation}
holds, yielding current conservation at $t=t_0$. Using the above in Eq. (\ref{NANMGauss}), we obtain $\dot{\Omega}^a(t=t_0)=0$. 
Since the relations $\Omega^a(t=t_0)=0$ and $\dot{\Omega}^a(t=t_0)=0$ are fulfilled, we obtain
\begin{eqnarray}
\nonumber
\Omega^a(t=t_0+\delta t_1)&=&\Omega^a(t=t_0)+\dot{\Omega}^a(t=t_0)\delta t_1+\cdots, \\
&=&0.
\end{eqnarray}

Using that $\Omega^a(t=t_0+\delta t_1)=0$, we observe that the Yang Mills equations are now valid at $t=t_0+\delta t_1$, which again, due to antisymmetry of the Maxwell tensor, imply current conservation at $t=t_0+\delta t_1$ and $\dot{\Omega}^a(t=t_0+\delta t_1)=0$ via Eq. (\ref{NANMGauss}).
The relations $\Omega^a(t=t_0+\delta t_1)=0$ and $\dot{\Omega}^a(t=t_0+\delta t_1)=0$ imply $\Omega^a(t=t_0+\delta t_1+\delta t_2)=0$. Iterating the previous process, always applying the dynamical equation (\ref{NANMGauss}) of the ENM, it follows that current conservation is valid for all time. Summarizing, the above analysis shows that in the case of the  non-Abelian Nambu model, the specific gauge structure of the theory allows us to  impose only the  Gauss  constraints as initial conditions, which necessarily yield current conservation for all time, in order to obtain  the corresponding  MGT.

\subsection{Linearized Gravity}
The main example in this paper corresponds to present the equivalence between the linearized Einstein gravity and one ENM. We start from the Fierz-Pauli Lagrange density  \cite{LFierz,ADMDirac,Lingrav2,Lingrav3,Lingrav4}
\begin{equation}
\mathcal{L}=\frac{1}{2}[\partial_\lambda h_{\mu\nu}\partial^\mu h^{\lambda\nu}-\partial_\lambda h \partial_\mu h^{\mu\lambda}]+\frac{1}{4}[\partial_\mu h \partial^\mu h-\partial_\lambda h_{\mu\nu}\partial^\lambda h^{\mu\nu}],
\end{equation}
where $h_{\mu\nu}$ is a second rank symmetric tensor, $h=\eta^{\mu\nu}h_{\mu\nu}$ with $\eta_{\mu\nu}=diag(-,+,+,+)$ being the metric tensor of the flat space. The non trivial gauge transformations in the above Lagrangian are $\delta h_{\mu\nu}=\partial_\mu \xi_\nu(x)+\partial_\nu \xi_\mu(x)$. Considering the action $\mathcal{S}=\int \,d^4x\, \mathcal{L}$, and after integration by parts, the Fierz-Pauli Lagrangian density can be rewritten as follows \cite{Lingrav2},
\begin{eqnarray}
\mathcal{L}&=&-\frac{1}{4}\dot{h}_{ii}\dot{h}_{jj}-\frac{1}{2}\partial_k h_{ii}\partial_kh_{00}+\frac{1}{4}\partial_ih_{jj}\partial_ih_{kk}+\frac{1}{4}\dot{h}_{ij}\dot{h}_{ij} \nonumber \\ && +\frac{1}{2}\partial_ih_{0j}\partial_ih_{0j}-\frac{1}{4}\partial_ih_{jk}\partial_i h_{jk} +\dot{h}_{ii}\partial_j h_{0j}-\frac{1}{2}\partial_i h_{kk}\partial_jh_{ij} \nonumber \\
&& +\frac{1}{2}\partial_i h_{00}\partial_j h_{ij} -\dot{h}_{ik}\partial_ih_{0k}-\frac{1}{2}\partial_ih_{0i}\partial_jh_{0j}+\frac{1}{2}\partial_ih_{jk}\partial_jh_{ik}.
\end{eqnarray}
where the latin indices stand for the pure space coordinates ($i,j=1,2,3$) and $\dot{f}\equiv\partial_0 f$. Introducing the momenta conjugate to $h_{\mu\nu}$ ($\mu,\nu=0,1,2,3$) 
\begin{equation}
p_{\mu\nu}=\frac{\delta \mathcal{L}}{\delta{\dot{h}_{\mu\nu}}},
\end{equation}
one can directly write the primary constraints
\begin{equation}
p_{0\nu}=0.
\label{graviconstr}
\end{equation}
The non-zero momenta are given by
\begin{equation}
p_{ij}=-\frac{1}{2}\delta_{ij}\dot{h}_{kk}+\frac{1}{2}\dot{h}_{ij}+\delta_{ij}\partial_k h_{k0}-\frac{1}{2}(\partial_ih_{0j}+\partial_jh_{0i}).
\end{equation}
The canonical Hamiltonian density is \cite{Lingrav2}
\begin{eqnarray}
\label{hamiltgravit}
\mathcal{H}_c&=&p^{ij}\dot{h}_{ij}-\mathcal{L}, \nonumber \\
&=&p_{ij}p_{ij}-\frac{1}{2}p_{kk}p_{ll}+\frac{1}{2}(\partial_ih_{kk}\partial_jh_{ij}-\partial_ih_{jk}\partial_jh_{ik})+\frac{1}{4}(\partial_ih_{jk}\partial_ih_{jk}-\partial_ih_{jj}\partial_ih_{kk}) \nonumber \\
&&-\frac{1}{2}h_{00}(\partial_i\partial_ih_{kk}-\partial_i\partial_jh_{ij})-2h_{0j}\partial_ip_{ij},
 \nonumber \\
&=& \mathcal{H}_F(h_{ij},p_{ij})-\frac{1}{2}h_{00}(\partial_i\partial_ih_{kk}-\partial_i\partial_jh_{ij})-2h_{0j}\partial_ip_{ij},
\end{eqnarray}
where
\begin{equation}
 \mathcal{H}_F(h_{ij},p_{ij})=p_{ij}p_{ij}-\frac{1}{2}p_{kk}p_{ll}+\frac{1}{2}(\partial_ih_{kk}\partial_jh_{ij}-\partial_ih_{jk}\partial_jh_{ik})+\frac{1}{4}(\partial_ih_{jk}\partial_ih_{jk}-\partial_ih_{jj}\partial_ih_{kk})\,.
\end{equation}
The time evolution of the primary constraints $\Omega^\mu=p^{0\mu}=0$ gives the secondary constraints $\Omega^4=\partial_i\partial_ih_{kk}-\partial_i\partial_jh_{ij}$ and $\Omega^{4+i}=\partial_jp_{ji}$. No tertiary constraints appear and the Dirac's method closes. 

It can be proved that all constraints $\Omega$'s have vanishing Poisson brackets among themselves \cite{ADMDirac,Lingrav2,Lingrav3}, so all of them are of first class. Also, the conditions $\{\Omega^4,\mathcal{H}_F\}\approx 0$ and $\{\Omega^{4+i},\mathcal{H}_F\}\approx 0$ are fulfilled \cite{ADMDirac,Lingrav2,Lingrav3}. The properties stated in this paragraph,  together with (\ref{graviconstr}) and (\ref{hamiltgravit}) fulfill the basic requirements (\ref{hamiltonianocanonic}), (\ref{condition2}) and (\ref{condition3}). Following Section \ref{gaugetheory}, we identify $\lambda_m\rightarrow  h_{00}, \, h_{0j}$ and $G_m\rightarrow \partial_i\partial_ih_{kk}-\partial_i\partial_jh_{ij}, \, \partial_jp_{ji}$.

\section{Summary and conclusions}

Spontaneous Lorentz symmetry breaking (SLSB) has attracted considerable
attention in recent years \cite{Colladay-Kostelecky,Colladay-Kostelecky2}, both from the
experimental and theoretical points of view. One of the rewards in
considering SLSB is the possibility of giving a dynamical setting to the
gauge principle by considering photons, gravitons and non-Abelian gauge
fields as the massless Nambu-Goldstone bosons arising from such a breaking 
\cite
{Nambu-Progr,Azatov-Chkareuli,Urru-Mont,JLCH1,JLCH2,JLCH3,NANM_ES_UR,JLCH4,NANM-Abel,emergent_gauge,emergent_gauge2}.
This approach can be codified under the name of different Nambu models. Usually  these models arise as the low energy limit of spontaneous Lorentz symmetry breaking (SLSB) in the so called Bumblebee models \cite{BUMBLEBEE4,BUMBLEBEE42,BUMBLEBEE43,BUMBLEBEE44,BUMBLEBEE45,bumblebee1,HERNASKI,Constri}. In practice Nambu models turn out to be described by 
the Lagrangian density of a gauge theory, the mother gauge theory (MGT), plus  a non-linear constraint, arising from the non-zero vacuum
expectation value of the corresponding gauge fields due to the SLSB. This constraint is not considered as an standard gauge fixing, but is explicitly solved and substituted in the gauge Lagrangian thus destroying gauge invariance. Some explicit calculations, like for example in Refs. \citen{Nambu-Progr,Azatov-Chkareuli,NANM-Abel},
have shown that, under some conditions, the violation of the
Lorentz symmetry and the lack of gauge invariance, introduced by the
non-linear constraint, become unobservable in such a way that the appearing GBs
can be interpreted as the gauge particles of the original MGT. 

A natural question posed by this approach in the realm of gauge theories, is to determine under which conditions the recovery of an arbitrary MGT theory from the corresponding  Nambu model, defined by a general constraint over the coordinates, becomes possible. We  refer to these theories as extended Nambu models (ENM), to differentiate them from the case where the constraint is treated as a standard gauge fixing term. At this level, the mechanism for generating the constraint is irrelevant and the case of SLSB is taken only as a motivation, which naturally  bring  this problem under consideration. The equivalence between gauge theories and ENMs is
not straightforward and one has to consider the following issues: (i)
because of the additional constraint, ENM models are not gauge invariant,
(ii) the number of degrees of freedom in the ENM is larger than that
of the  MGT, and (iii) the equations of motion of the two theories do not
match. In this way, it becomes clear that additional requirements have to be
imposed upon the ENM in order to recover the original
MGT. The strategy we follow
is a generalization of the non-perturbative Hamiltonian analysis developed
for the abelian Nambu model in Ref. \citen{Urru-Mont} and for the non-abelian
Nambu model in Ref. \citen{NANM_ES_UR}.

In Section \ref{gaugetheory}, we define the class of MGT under consideration
by the requirements given in Eqs. (\ref{hamiltonianocanonic}), (\ref
{condition2}) and (\ref{condition3}), starting from a Lagrangian $
L(\theta_{l},\lambda _{A},\;\dot{\theta}_{l})$, where $l=1,2,...,n$  and $
A=1,2,...,K$. We also perform the Dirac method to calculate the Hamiltonian (\ref{hamiltonianofinal}) 
and the canonical algebra (\ref{algebrafinal}) of the MGT. Subsequently, the corresponding ENM is
introduced in Section \ref{NGmodel}, by having the same Lagrangian as the
MGT plus a constraint $F(\theta_{l},\lambda _{A})=0$ among the
coordinates of the MGT. We solve the nonlinear constraint (\ref{constriccion}) 
in two generic different ways, and we identify the canonical coordinates
and momenta of the ENM, together with their canonical algebra and extended
Hamiltonian. At this level, only second class constraints arise, reflecting
the fact that the ENM is a theory without gauge invariance. Since both
theories arise from the same Lagrangian, it is possible to rewrite the
standard canonical variables of the MGT in terms of the canonical variables
of the ENM. In this way, using the canonical algebra of the ENM, we show
that the canonical algebra of the\ MGT is recovered. \ref
{Apendice1} and \ref{Apendice2}  include the calculation of the required PBs that prove the
previous statement. The second class constraints in the ENM are
further imposed strongly, by introducing the corresponding Dirac brackets (DBs), in order to
eliminate the canonical variables $\lambda _{A},\;\bar{\Pi}_{B}^{\lambda }$.
The DBs of the remaining variables are also calculated yielding no
modifications with respect to the original PBs. The final extended
Hamiltonian for the ENM, rewritten in terms of the canonical variables of
the MGT, has the same form as the Hamiltonian corresponding to the MGT,
except that the Gauss functions $G_{A}$  do not appear as first class constraints, as it should be in the
MGT. This is because their coefficients in the extended Hamiltonian are not
arbitrary functions, but specific functions of the coordinates of the ENM.
In order to remedy this issue we calculate the time evolution of the
functions $G_{A}$, according to the ENM dynamics, and find that demanding  
generalized current conservation {$(DJ)_A=0$} for all time, together
$G_{A}=0$ as initial 
conditions yields $G_{A}(t)=0 $ for all time. This
allows us to include the quantities $G_{A}$ as first class constraints in the extended
Hamiltonian through arbitrary functions $N_{A}$ adding the term $N_{A}G_{A}$
to the ENM Hamiltonian. In this way, the Hamiltonian describing the ENM
becomes the same as the Hamiltonian of the  MGT. It is important to recall that the generalized current conservation
condition $(DJ)_A=0$ follows from the gauge invariance generated by the Gauss law constraints $G_A\approx 0$ in the Hamiltonian action of the MGT. We prove this statement in the \ref{CURCONS1}. 

Summarizing, the correct statement is that gauge invariance is indeed
dynamically recovered in the ENM provided we impose current conservation for all times and the Gauss laws only as initial
conditions. Let us recall that the canonical algebra of the MGT
has been already recovered from that of the ENM. In this way, the complete
equivalence of the MGT with the ENM plus those suitable conditions
is obtained.  We emphasize that such equivalence has been proved for an arbitrary gauge fixing in the MGT and that it is completely unrelated to the specific constraint which defines the ENM.

Section \ref{examples} presents some particular cases of MGTs in which our
general result can be directly applied to construct arbitrary ENMs, 
from where the MGT can be ultimately recovered after imposing
the appropriate conditions previously stated in our general analysis.
Electrodynamics, Yang Mills theories and linearized gravity are typical
examples described previously in the literature, where the constraint $
F(\theta_{l},\lambda _{A})=0$ arises from spontaneous Lorentz symmetry breaking.

Since on one hand the standard Nambu models make explicit reference to spontaneous Lorentz symmetry breaking
and on the other hand they constitute particular cases of the ENM we have
considered; our results confirm and clarify the statement that, under the
imposition of current conservation for all time together with the Gauss constraints 
as initial conditions, Lorentz invariance violation in such Nambu models is physically
unobservable.

\section*{Acknowledgements}

 L. F. U. has been partially  supported by the project CONACyT \# 237503. C. A. E. and L. F. U. acknowledge support from the project UNAM (Direcci\'on General de Asuntos del Personal Acad\'emico) \# IN104815. C. A. E. is supported by  the CONACyT Postdoctoral Grant No. 234745.

\appendix

\section{The generalized current conservation}
\label{CURCONS1}

{We generalize to the Hamiltonian formulation the general idea that current conservation is a consequence
of the invariance of the action under the gauge transformations, which in this case are  generated by the Gauss constraints. 
The simplest example of this relation arises in the Lagrangian formulation of electrodynamics with the action}
\begin{equation}
S=\int d^4x \, \left(-\frac{1}{4} F^{\mu\nu} F_{\mu\nu}-J^\mu A_\mu\right).
\end{equation}
{Demanding $\delta S=0$ under the gauge transformations $\delta A_\mu= \partial_\mu \alpha$, with parameter $\alpha$, yields}
\begin{equation}
0=\delta S= -\int d^4x\, \alpha \,(\partial_\mu J^\mu), \quad \longrightarrow \quad \partial_\mu J^\mu=0, 
\end{equation}
after integrating by parts.
In our case we start from the MGT described by the extended Hamiltonian density (\ref{hamiltonianofinal})
\begin{equation}
{\cal H}_E= {\cal H}_F+ \Theta_A (G^0_A+ J_A) + \theta_l J_l,
\end{equation}
where we have fixed the constraints $\lambda_A= \Theta_A, \, \Pi^\lambda_A=0$ such that the corresponding canonical variables $\theta_k, \, \Pi^\theta_l$ satisfy the standard Dirac brackets (\ref{algebrafinal}). Here $\Theta_A$ play the r\^ole of Lagrange multipliers.

The Hamiltonian action can be written in a compact way as 
\begin{equation}
S=\int dt\left[ \Pi _{k}^{\theta }\;\dot{\theta}_{k}-\left( {H}%
_{F}+\Theta_{A}(G_{A}^{0}+J_{A}\right) +\theta _{l}J_{l})\right],
\label{HAMACTION} 
\end{equation}
with the convention that the contraction of the indices $k$ and $A$ include an integral over the respective coordinates. In other words we denote
\begin{equation}
P_kQ_k=\int d^3x\, P_k(t,{\mathbf x})Q_k(t,{\mathbf x}), \quad R_AS_A=\int d^3x\, R_A(t,{\mathbf x})S_A(t,{\mathbf x}).
\end{equation}
Also, $H_F$ is the Hamiltonian corresponding to the Hamiltonian density ${\mathcal H}_F$ and we identify $G_A=G_A^0 +J_A$ as the remaining first class constraints. Let us 
recall that $J_{A},\, J_{l}$ are external currents independent of the canonical variables. 

The following calculation is an extension of the discussion related to the gauge invariance of the action in Chapter 3 of Ref. \citen{HT} and we heavily rely upon the results included there. Let us  consider the  change of the action under the gauge transformations generated by the Gauss constraints $G_A$. The general transformation of any function of the canonical variables is
\begin{equation}
\delta F=\;\alpha _{B}\left\{ F,G_{B}\right\} =\alpha _{B}\left\{
F,G_{B}^{0}\right\}, 
\label{CHANGE}
\end{equation}
where $\alpha_A(x)$ are the gauge parameters. In particular, one can show \cite{HT}
\begin{equation}
\delta \left( \Pi _{k}^{\theta }\;\dot{\theta}_{k}\right) =\frac{d}{dt}\left[
\alpha _{A}\left( \frac{\partial G_{A}^{0}}{\partial \Pi _{k}^{\theta }}\Pi
_{k}^{\theta }-G_{A}^{0}\right) \right] +\frac{\partial \alpha _{A}}{%
\partial t}G_{A}^{0}.
\end{equation}
Equation (\ref{CHANGE}) also leads to 
\begin{eqnarray}
&&\delta H_{F}=\alpha_{A}\left\{ H_{F},G_{A}^{0}\right\} =-\alpha
_{A}C_{AB}G_{B}^{0}, \nonumber \\
&&\delta \theta _{l}=\;\alpha _{B}\left\{ \theta _{l},G_{B}\right\} =\alpha
_{A}\left\{ \theta _{l},G_{A}^{0}\right\}, \nonumber \\
&&\delta G_{A}^{0}=\;\alpha _{B}\left\{ G^0_{A},G_{B}\right\}
=\alpha _{B}\left\{ G_{A}^{0},G_{B}^{0}\right\} =C_{ABC}\alpha _{B}G_{C}^{0}.
\end{eqnarray}
The next step is to establish the transformation law for the Lagrange multipliers. According to Ref. \citen{HT} this is given by 
\begin{equation}
\delta \Theta_{A}=\frac{\partial \alpha ^{A}}{\partial t}+\alpha
_{B} \Theta_{C}C_{BCA}+\alpha _{B}C_{BA}.
\end{equation}
In this way we obtain
\begin{eqnarray}
\delta S &=&\int dt \left[ \frac{\partial \alpha _{A}}{\partial t}G_{A}^{0}
+ \alpha_{A}C_{AB}G_{B}^{0}- \alpha_{B}C_{BA}G_{A}^{0} \right. \nonumber \\
&& \left. -\frac{\partial
\alpha _{A}}{\partial t}G_{A}^{0} - [C_{BAC}+C_{ABC}] \Theta_{A}\alpha
_{B}G_{C}^{0}\right] \nonumber \\
&& - \int dt\left[ \left( \frac{\partial \alpha _{A}}{\partial t}%
+\Theta_{C}\alpha _{B}C_{BCA}+\alpha _{B}C_{BA}\right) J_{A}+\delta \theta
_{l}J_{l}\right], 
\label{FINALCONS}
\end{eqnarray}
where we have separated the term dependent on the currents in the third line of the above equation. It is interesting to observe that the cancellation in the second line of Eq. (\ref{FINALCONS})  depends only on the antisymmetry of $C_{ABC}$ on the first two index, as required by their definition. This cancellation is the statement of the gauge invariance of the MGT in the absence of external currents. 

Imposing $\delta S=0$ in the general case we are left with
\begin{equation}
\frac{\partial J_{A}(x)}{\partial t}-C_{ABC}\lambda_{B}J_{C}(x)-C_{AB}J_{B}(x)+ \int d^3y\left\{
 \, G_{A}^{0}(x),\theta_{l}(y)\right\} J_{l}(y)=0,
\label{FINCURCONS1}
\end{equation}
which is our generalized statement of current conservation. In the above equation, which reproduces Eqs. (\ref{DEFDJA}) and (\ref{currentcons}),  we have used $\Theta_A=\lambda_A$ and we have restated the coordinate dependence. We find it very remarkable that the above statement of current conservation can be obtained without making any reference to the detailed structure of the constraints $G_A$.

\section{A Mechanical Model: detailed calculation}
\label{Mechanics1a}

In this Appendix we explicitly show how the equivalence between the MGT and the ENM is carried out for the mechanical model defined in Eq.(\ref{MECMOD}) of Section \ref{examples}, which is a theory with first class constraints and non trivial  gauge transformations that defines the mother gauge theory (MGT). The Dirac algorithm is performed and the canonical algebra, together with the Hamiltonian describing the dynamics of this MGT are presented. The next step is to build the ENM and to prove the equivalence with the MGT, showing that both the Hamiltonian and the algebra of this ENM correspond to those of the MGT, after suitable conditions are imposed to the ENM.

\subsection{The Gauge theory}

The MGT is defined by the Lagrangian
\begin{equation}
{L}=\frac{1}{2}(\dot{x}^{2}+\dot{y}^{2})-\lambda (x\dot{y}-y\dot{x})+
\frac{1}{2}\lambda ^{2}(x^{2}+y^{2})-V(x^{2}+y^{2}).
\label{MGTL}
\end{equation}
It can be easily proved that, the canonical Hamiltonian is given by 
\begin{eqnarray}
\nonumber
{H}_{c} &=& \dot{x}\Pi_x+\dot{y}\Pi_y-{{L}}, \\ \nonumber
&=&\frac{\Pi _{x}^{2}}{2}+\frac{\Pi _{y}^{2}}{2}
+V(x^{2}+y^{2})+\lambda (x\Pi _{y}-y\Pi _{x}), \\
&=&{H}_{F}(x,y,\Pi _{x},\Pi _{y})+\lambda (x\Pi _{y}-y\Pi _{x}), 
\label{HamilMGT1a}
\end{eqnarray}
where 
\begin{equation}
\Pi_\lambda=\frac{\partial \mathcal{L}}{\partial\dot{\lambda}}=0,\quad\quad
 \Pi _{x}=\frac{\partial \mathcal{L}}{
\partial \dot{x}}=\dot{x}+\lambda y,\quad \quad  \Pi _{y}=\frac{\partial \mathcal{L}}{
\partial \dot{y}}=\dot{y}-\lambda x,
\end{equation}
\begin{equation}
{H}_{F}=\frac{\Pi _{x}^{2}}{2}+\frac{\Pi _{y}^{2}}{2}
+V(x^{2}+y^{2}).
\label{canmomegaug}
\end{equation}
The canonical algebra is given by the non-zero Poisson brackets 
\begin{equation}
\{x,\Pi_x\}=\{y,\Pi_y\}=\{\lambda,\Pi_\lambda\}=1.
\end{equation}
The model only has the primary constraint $\Phi _{1}=\Pi _{\lambda }\approx0$, which leads to the secondary constraint $\Phi _{2}=x\Pi _{y}-y\Pi _{x}\approx0$. They are first class constraints, which implies that the model has one DOF. This agrees with the counting in Eq. (\ref{dofgauge}). Now we construct the extended Hamiltonian
\begin{equation}
{H}_{E} = {H}_{c}+ \beta\,\Pi_\lambda,
\end{equation}
where $\beta$ is a Lagrange multiplier.  In order to eliminate $\lambda$ and $\Pi_\lambda$ we add the gauge condition $\Phi_3=\lambda-\Theta\approx0$, where $\Theta$ is an arbitrary function to be consistently determined after the remaining first class constraint $\Phi_2$ is fixed. The constraints $\Phi_1$ and $\Phi_3$ become second class and we can introduce the Dirac's brackets to describe the dynamics. The matrix constructed with  $\Phi_1$ and $\Phi_3$ is given by
\[ M_{ij} = \left( 
\begin{array}{cc}
0 & -1 \\
1 & 0  
\end{array} \right),\] 
with the inverse 
\[ (M^{-1})_{ij} = \left( 
\begin{array}{cc}
0 & 1 \\
-1 & 0  
\end{array} \right),\] 
Using the above matrix, we calculate the Dirac's brackets for the remaining variables with the result
\begin{equation}
\{x,x\}_D=\{y,y\}_D=0,\,\, 
\{\Pi_x,\Pi_x\}_D=\{\Pi_y,\Pi_y\}_D=0,\,\,
\{x,\Pi_y\}_D=\{y,\Pi_x\}_D=0,
\label{algex1}
\end{equation}
\begin{equation}
\{x,\Pi_x\}_D=\{y,\Pi_y\}_D=1.
\label{algex2}
\end{equation}
Fixing strongly the constraints $\Phi_1=\Pi_\lambda=0$ and $\Phi_3=\lambda-\Theta=0$, together with the introduction of the Dirac's brackets, the Hamiltonian becomes
\begin{eqnarray}
\nonumber
{H}_{E} &=&\frac{\Pi _{x}^{2}}{2}+\frac{\Pi _{y}^{2}}{2}
+V(x^{2}+y^{2})+\Theta (x\Pi _{y}-y\Pi _{x}), \\
\label{HamilEx2}
&=&{H}_{F}(x,y,\Pi _{x},\Pi _{y})+\Theta (x\Pi _{y}-y\Pi _{x}).
\end{eqnarray}
At this point, the dynamics of this theory is determined by the extended Hamiltonian ${H}_{E}$ given in Eq. (\ref{HamilEx2}), together with the algebra (\ref{algex1}-\ref{algex2}). In the next Subsection we are going to define an extended Nambu model using the Lagrangian (\ref{MGTL}), plus a relation between the coordinates. The goal will be to recover the extended Hamiltonian ${H}_{E}$ and the algebra (\ref{algex1}-\ref{algex2}) using the dynamics of the ENM.

\subsection{The extended Nambu model}
According to Section \ref{NGmodel}, we can construct a Nambu model using the above Lagrangian plus a suitable relation among the coordinates $(x,y,\lambda)$. We illustrate the two generic cases: a) solving the coordinate $x$ and b) solving the coordinate $\lambda$. We choose a particular function in order to show explicitly the equivalence.

\subsubsection{Solving for the coordinate $x$}
In this case we define the extended Nambu model by the Lagrangian
\begin{equation}
{L}=\frac{1}{2}(\dot{x}^{2}+\dot{y}^{2})-\lambda (x\dot{y}-y\dot{x})+
\frac{1}{2}\lambda ^{2}(x^{2}+y^{2})-V(x^{2}+y^{2}),
\label{LagranC1}
\end{equation}
plus the relation
\begin{equation}
x=x(y,\lambda)=\frac{1}{2}(y^2+\lambda^2),
\label{ConsC1}
\end{equation}
which implies 
\begin{equation}
\dot{x}=y\dot{y}+\lambda\dot{\lambda}.
\label{Constric1a}
\end{equation}
When Eq. (\ref{Constric1a}) is substituted into the Lagrangian (\ref{LagranC1}), we obtain the Lagrangian 
\begin{equation}\tilde{ L}(y, \lambda)={ L}(x=x(y, \lambda), y, \lambda)
\end{equation} 
for the ENM, where we also substitute ${\dot x}=y {\dot y} + \lambda {\dot \lambda}$. The canonical momenta can be obtained making use of the chain rule and they are given by
\begin{equation}
\tilde{\Pi}_y= \frac{\partial \tilde{ L}}{\partial \dot{y}}=\dot{y}-\lambda x +(\dot{x}+\lambda y)y, \qquad \tilde{\Pi}_\lambda= \frac{\partial \tilde{ L}}{\partial \dot{\lambda}} =(\dot{x}+\lambda y )\lambda.
\label{canmometa1}
\end{equation}
We emphasize that  $x$ and $\dot{x}$ in the above expressions are just labels for the combinations (\ref{ConsC1}) and (\ref{Constric1a}), respectively. In other words, $x$ is not a coordinate of the ENM. It can be shown that the  relations (\ref{canmometa1}) can be solved for the velocities $\dot{y}$ and $\dot{\lambda}$ in terms of the momenta, such that there are no constraints in the theory; therefore, the number of DOF is two, which agrees with Eq. (\ref{dofx1a}).
 
The canonical algebra between the coordinates $y$, $\lambda$ and their momenta is given by the non-zero Poisson brackets
\begin{equation}
\{y,\tilde{\Pi}_y\}=1,\quad\quad\quad\{\lambda,\tilde{\Pi}_\lambda\}=1.
\label{algxc1}
\end{equation}
 We note that we can rewrite the canonical momenta (\ref{canmometa1}) in terms of the canonical momenta of the gauge theory (\ref{canmomegaug}) as
\begin{equation}
\tilde{\Pi}_y=\Pi_y+\Pi_x y,\quad\quad\quad \tilde{\Pi}_\lambda=\Pi_x \lambda,
\label{moment1ab}
\end{equation}
where the definitions
\begin{equation}
\Pi_x=\dot{x}+\lambda y,\quad\quad  \Pi_y=\dot{y}-\lambda x,
\end{equation} 
are, again,  just labels to specify a particular combination of the variables $(y,\lambda)$, their velocities and their momenta.
 
Solving the expressions in Eq. (\ref{moment1ab}) in favor of $\Pi_x$ and $\Pi_y$ we have
\begin{equation}
\Pi_x=\frac{\tilde{\Pi}_\lambda}{\lambda},\quad\quad\quad \Pi_y= \tilde{\Pi}_y-\frac{\tilde{\Pi}_\lambda}{\lambda} y.
\end{equation}
{Next we calculate} the Poisson brackets among the following  quantities: $x=x(y,\lambda), \,\, y, \,\, \Pi_x=\Pi_x(y,\lambda,\tilde{\Pi}_y,\tilde{\Pi}_\lambda)$ and $ \Pi_y=\Pi_y(y,\lambda,\tilde{\Pi}_y,\tilde{\Pi}_\lambda)\}$, using the canonical algebra of the ENM in Eq. (\ref{algxc1}).  We find that the only non-zero Poisson brackets are 
\begin{equation}
\{x,\Pi_x\}=1,\quad\quad\quad\{y,\Pi_y\}=1.
\label{CANAENM}
\end{equation}
The canonical Hamiltonian of the ENM can be computed as
\begin{eqnarray}
{H}_c^{ENM}&=& \dot{y}\tilde{\Pi}_y+\dot{\lambda}\tilde{\Pi}_\lambda- \tilde{ L}, \nonumber \\
&=&  \dot{y}(\Pi_y+\Pi_x y)+\dot{\lambda} \Pi_x \lambda -\tilde{ L}, \nonumber \\
&=&  \dot{y}\Pi_y+(y \dot{y}+\lambda\dot{\lambda})\Pi_x-\tilde{ L}, \nonumber \\
&=&  \dot{y}\Pi_y+\dot{x}\Pi_x-{L}.
\end{eqnarray}
In the last line of the above equation we have undone the substitution (\ref{ConsC1}) in $\tilde{L}$ and we recover ${L}(x,y,\lambda)$. After substituting the velocities $\dot{x}$ and $\dot{y}$ in terms of the momenta  $\Pi_x$ and $\Pi_y$, according to Eq.(\ref{canmomegaug}), we can write
\begin{equation}
{H}_c^{ENM}=\frac{\Pi _{x}^{2}}{2}+\frac{\Pi _{y}^{2}}{2}
+V(x^{2}+y^{2})+\lambda (x\Pi _{y}-y\Pi _{x}).
\label{HamilC11}
\end{equation}
When going back to the reduced MGT variables $x$ and $y$ we  recall that the relation (\ref{ConsC1}) yields $\lambda=\lambda(x,y)=\sqrt{2x- y^2}$ in our case. 
At this point we have recovered in Eq.(\ref{CANAENM}) the canonical algebra of the MGT given in Eqs (\ref{algex1}-\ref{algex2}). Also, the canonical Hamiltonian of the ENM in Eq.(\ref{HamilC11}) has the same form of the extended Hamiltonian of the MGT in Eq. (\ref{HamilEx2}). The only difference arises from the fact that in this case the quantity $G\equiv(x\Pi _{y}-y\Pi _{x})$ is not a constraint, which is reflected in that the factor $\lambda$ is not an arbitrary function. 

To deal with this issue we study the time evolution of the quantity $G$ in the ENM. The dynamics of the ENM leads to  
\begin{equation}
\dot{G}=\{ (x\Pi _{y}-y\Pi _{x}),{H}_c^{ENM}\}=\{(x\Pi _{y}-y\Pi _{x}),\lambda(x,y)\}G=\frac{y}{\lambda}(1+x)\, G.
\end{equation}
The evaluation of the previous bracket is performed in an easier way by using the algebra of the MGT, which we have previously proved  that can be derived from the ENM algebra. The above equation shows that if we demand that $G(t=0)=0$, we obtain that $\dot{G}=0$ as well at t=0. This proves that the relation $G(t)=0$ will hold for all time.   
Therefore, we can include the quantity $G\equiv(x\Pi _{y}-y\Pi _{x})$ as a constraint in the Hamiltonian (\ref{HamilC11}), through an arbitrary function $N$, by adding $NG$ and redefining $\lambda+N=\Theta$. We obtain 
\begin{equation}
{H}_{c} ={H}_{F}(x,y,\Pi _{x},\Pi _{y})+\Theta (x\Pi _{y}-y\Pi _{x}), 
\end{equation}
where now $\Theta$ is an arbitrary function. In this way we recover the extended Hamiltonian (\ref{HamilEx2}) together the algebra (\ref{algex1}-\ref{algex2}) of the MGT, thus proving  the equivalence between the ENM and the MGT for this case, once  $G=0$ has been imposed as an initial condition.  

\subsubsection{Solving for the coordinate $\lambda$}

Now we define the extended Nambu model by the Lagrangian
\begin{equation}
{L}=\frac{1}{2}(\dot{x}^{2}+\dot{y}^{2})-\lambda (x\dot{y}-y\dot{x})+
\frac{1}{2}\lambda ^{2}(x^{2}+y^{2})-V(x^{2}+y^{2}),
\label{LagranC2}
\end{equation}
plus the relation
\begin{equation}
\lambda=\lambda(x,y)=x^2+y^2.
\label{ConsC2}
\end{equation}
Since the relation (\ref{ConsC2}) does not modify the velocities of the Lagrangian (\ref{LagranC2}), because $\dot{\lambda}$ is not present there, the canonical momenta $\Pi_x$ and $\Pi_y$ coincide with those of the MGT theory and satisfy the canonical algebra given by the non-zero Poisson brackets
\begin{equation}
\{x,\Pi_x\}=\{y,\Pi_y\}=1.
\end{equation}
As anticipated by Eqs. (\ref{constr123a}) and (\ref{dof1ab}) there are no constraints and the number of DOF in this model is two. The canonical Hamiltonian is obtained in the same manner as in the MGT, leading to  
\begin{eqnarray}
\nonumber
{H}_{c}^{ENM} &=&\frac{\Pi _{x}^{2}}{2}+\frac{\Pi _{y}^{2}}{2}
+V(x^{2}+y^{2})+\lambda(x,y) (x\Pi _{y}-y\Pi _{x}), \\
\label{HamilC2}
&=&{H}_{F}(x,y,\Pi _{x},\Pi _{y})+\lambda(x,y) (x\Pi _{y}-y\Pi _{x}), 
\end{eqnarray}
where $\lambda(x,y)$ is not an arbitrary function, but it is given by Eq. (\ref{ConsC2}). In this model, the quantity $G\equiv(x\Pi _{y}-y\Pi _{x})$ is not a constraint; however, it is a conserved quantity, which follows from the bracket
\begin{equation}
\dot{G}=\{ (x\Pi _{y}-y\Pi _{x}),{H}_{c}\}=0.
\end{equation}
This show that if we demand $G(t=0)=0$, then $G(t)=0$ for all times. Therefore, we can include the quantity $G\equiv(x\Pi _{y}-y\Pi _{x})$ as a constraint in the Hamiltonian (\ref{HamilC2}), through an arbitrary function $N$, by adding $NG$ and redefining $\lambda(x,y)+N=\Theta$. We obtain 
\begin{equation}
{H}_{c} ={H}_{F}(x,y,\Pi _{x},\Pi _{y})+\Theta (x\Pi _{y}-y\Pi _{x}), 
\end{equation}
where now $\Theta$ is an arbitrary function. Therefore, we recover the Hamiltonian (\ref{HamilEx2}) together the algebra (\ref{algex1}-\ref{algex2}) of the MGT.

\section{The algebra resulting  when solving for the coordinate $\theta_{1}$  in the ENM
constraint}
\label{Apendice1}

In this Appendix we show that the algebra for the canonical
variables $\theta_{i}$ and $\Pi _{j}^{\theta}$ in the\ MGT, given in Eq. (\ref
{algebrafinal}), is recovered from the algebra of canonical variables
corresponding to the ENM given in Subsection \ref{x1}. 
The algebra of the second class constraints of the
ENM is also calculated.

Here the canonical variables of the ENM ($\theta_{\bar{l}},\lambda_{A},\bar{\Pi}
_{\bar{l}}^{\theta},\bar{\Pi}_{B}^{\lambda }$) have the non-zero PBs 
\begin{equation}
\{\theta_{\bar{\imath}},\bar{\Pi}_{\bar{j}}^{\theta}\}=\delta _{{\bar{\imath}}{\bar{j}},
}\quad \quad \quad \quad \{\lambda _{A},\bar{\Pi}_{B}^{\lambda }\}=\delta
_{AB}.\quad \quad \quad (\bar{\imath},\bar{j}=2,3,...,n).
\label{commutation_constric2}
\end{equation}
The transformation from canonical variables of the ENM to those of the MGT is given by 
\begin{equation}
\theta_{1}=f(\theta_{\bar{l}},\lambda _{A}),\quad   \theta_{\bar{l}}=\theta_{
\bar{l}},\quad  \Pi _{1}^{\theta}=\frac{\bar{\Pi}_{1}^{\lambda }}{
f_{\lambda _{1}}},\quad   \Pi _{\bar{l}}^{\theta}=\bar{\Pi}_{\bar{l}
}^{\theta}-\frac{\bar{\Pi}_{1}^{\lambda }}{f_{\lambda _{1}}}f_{\theta_{\bar{l}}},\quad
  (\bar{l}=2,3,...,n).  \label{transformation2}
\end{equation}
Our goal is to calculated the algebra among the variables $\theta_{i},\Pi _{j}$
of the MGT in terms of the canonical algebra (\ref
{commutation_constric2}) of the ENM.
\subsection{The $\theta_i-\theta_j$ sector}
\begin{equation}
\{\theta_{1},\theta_{1}\}=\{f(\theta_{\bar{\imath}},\lambda _{A}),f(\theta_{\bar{j}},\lambda
_{A})\}=0.
\end{equation}
\begin{equation}
\{\theta_{1},\theta_{l}\}=\{f(\theta_{\bar{n}},\lambda _{A}),\theta_{\bar{l}}\}=0.
\end{equation}
\begin{equation}
\{\theta_{\bar{l}},\theta_{\bar{m}}\}=0.
\end{equation}
\subsection{The $\Pi^\theta_i-\Pi^\theta_j$ sector}
\begin{equation}
\{\Pi _{1}^{\theta},\Pi _{1}^{\theta}\}=\bigg\{\frac{\bar{\Pi}_{1}^{\lambda }}{
f_{\lambda _{1}}},\frac{\bar{\Pi}_{1}^{\lambda }}{f_{\lambda _{1}}}\bigg\}=0.
\end{equation}
\begin{eqnarray}
\{\Pi _{1}^{\theta},\Pi _{\bar{l}}^{\theta}\} &=&\bigg\{\frac{\bar{\Pi}_{1}^{\lambda }
}{f_{\lambda _{1}}},\bar{\Pi}_{\bar{l}}^{\theta}-\frac{\bar{\Pi}_{1}^{\lambda }}{
f_{\lambda _{1}}}f_{\theta_{\bar{l}}}\bigg\}=\bigg\{\frac{\bar{\Pi}_{1}^{\lambda }
}{f_{\lambda _{1}}},\bar{\Pi}_{\bar{l}}^{\theta}\bigg\}-\frac{\bar{\Pi}
_{1}^{\lambda }}{f_{\lambda _{1}}}\bigg\{\frac{\bar{\Pi}_{1}^{\lambda }}{
f_{\lambda _{1}}},f_{\theta_{\bar{l}}}\bigg\},  \notag \\
&=&\bar{\Pi}_{1}^{\lambda }\bigg\{\frac{1}{f_{\lambda _{1}}},\bar{\Pi}_{\bar{
l}}^{\theta}\bigg\}-\frac{\bar{\Pi}_{1}^{\lambda }}{f_{\lambda _{1}}f_{\lambda
_{1}}}\bigg\{\bar{\Pi}_{1}^{\lambda },f_{\theta_{\bar{l}}}\bigg\},  \notag \\
&=&-\frac{\bar{\Pi}_{1}^{\lambda }}{f_{\lambda _{1}}^{2}}f_{{\lambda _{1}}\theta_{
\bar{l}}}+\frac{\bar{\Pi}_{1}^{\lambda }}{f_{\lambda _{1}}^{2}}f_{\theta_{\bar{l}
}\lambda _{1}}=0.
\end{eqnarray}
\begin{eqnarray}
\{\Pi _{\bar{l}}^{\theta},\Pi _{\bar{m}}^{\theta}\} &=&\bigg\{\bar{\Pi}_{\bar{l}}^{\theta}-
\frac{\bar{\Pi}_{1}^{\lambda }}{f_{\lambda _{1}}}f_{\theta_{\bar{l}}},\bar{\Pi}_{
\bar{m}}^{\theta}-\frac{\bar{\Pi}_{1}^{\lambda }}{f_{\lambda _{1}}}f_{\theta_{\bar{m}}}
\bigg\},  \notag \\
&=&-\bigg\{\bar{\Pi}_{\bar{l}}^{\theta},\frac{f_{\theta_{\bar{m}}}}{f_{\lambda _{1}}}
\bigg\}\bar{\Pi}_{1}^{\lambda }+\bigg\{\bar{\Pi}_{\bar{m}}^{\theta},\frac{f_{\theta_{
\bar{l}}}}{f_{\lambda _{1}}}\bigg\}\bar{\Pi}_{1}^{\lambda }  \notag \\
&&+\bigg\{\bar{\Pi}_{1}^{\lambda },\frac{f_{\theta_{\bar{m}}}}{f_{\lambda _{1}}}
\bigg\}\bar{\Pi}_{1}^{\lambda }\frac{f_{\theta_{\bar{l}}}}{f_{\lambda _{1}}}-
\bigg\{\bar{\Pi}_{1}^{\lambda },\frac{f_{\theta_{\bar{l}}}}{f_{\lambda _{1}}}
\bigg\}\bar{\Pi}_{1}^{\lambda }\frac{f_{\theta_{\bar{m}}}}{f_{\lambda _{1}}}, 
\notag \\
&=&\bigg(\frac{f_{\theta_{\bar{m}}\theta_{\bar{l}}}}{f_{\lambda _{1}}}-\frac{f_{\theta_{
\bar{m}}}f_{\lambda _{1}\theta_{\bar{l}}}}{(f_{\lambda _{1}})^{2}}\bigg)\bar{\Pi}
_{1}^{\lambda }-\bigg(\frac{f_{\theta_{\bar{l}}\theta_{\bar{m}}}}{f_{\lambda _{1}}}-
\frac{f_{\theta_{\bar{l}}}f_{\lambda _{1}\theta_{\bar{m}}}}{(f_{\lambda _{1}})^{2}}
\bigg)\bar{\Pi}_{1}^{\lambda }  \notag \\
&&-\bigg(\frac{f_{\theta_{\bar{m}}\lambda _{1}}}{f_{\lambda _{1}}}-\frac{f_{\theta_{
\bar{m}}}f_{\lambda _{1}\lambda _{1}}}{(f_{\lambda _{1}})^{2}}\bigg)\bar{\Pi}
_{1}^{\lambda }\frac{f_{\theta_{\bar{l}}}}{f_{\lambda _{1}}}+\bigg(\frac{f_{\theta_{
\bar{l}}\lambda _{1}}}{f_{\lambda _{1}}}-\frac{f_{\theta_{\bar{l}}}f_{\lambda
_{1}\lambda _{1}}}{(f_{\lambda _{1}})^{2}}\bigg)\bar{\Pi}_{1}^{\lambda }
\frac{f_{\theta_{\bar{m}}}}{f_{\lambda _{1}}},  \notag \\
&=&-\bigg(\frac{f_{\theta_{\bar{m}}}f_{\lambda _{1}\theta_{\bar{l}}}}{(f_{\lambda
_{1}})^{2}}\bigg)\bar{\Pi}_{1}^{\lambda }+\bigg(\frac{f_{\theta_{\bar{l}
}}f_{\lambda _{1}\theta_{\bar{m}}}}{(f_{\lambda _{1}})^{2}}\bigg)\bar{\Pi}
_{1}^{\lambda }  \notag \\
&&-\bigg(\frac{f_{\theta_{\bar{m}}\lambda _{1}}}{f_{\lambda _{1}}}\bigg)\bar{\Pi}
_{1}^{\lambda }\frac{f_{\theta_{\bar{l}}}}{f_{\lambda _{1}}}+\bigg(\frac{f_{\theta_{
\bar{l}}\lambda _{1}}}{f_{\lambda _{1}}}\bigg)\bar{\Pi}_{1}^{\lambda }\frac{
f_{\theta_{\bar{m}}}}{f_{\lambda _{1}}}=0.
\end{eqnarray}
\subsection{The $\theta_i-\Pi^\theta_j$ sector}
\begin{equation}
\{\theta_{1},\Pi _{1}^{\theta}\}=\bigg\{f(\theta_{\bar{l}},\lambda _{A}),\frac{\bar{\Pi}
_{1}^{\lambda }}{f_{\lambda _{1}}}\bigg\}=\frac{f_{\lambda _{1}}}{f_{\lambda
_{1}}}=1.
\end{equation}
\begin{eqnarray}
\{\theta_{1},\Pi _{\bar{l}}^{\theta}\} &=&\{\theta_{1},\bar{\Pi}_{\bar{l}}^{\theta}-\Pi
_{1}^{\theta}f_{\theta_{\bar{l}}}\}, \notag \\
&=&\{\theta_{1},\bar{\Pi}_{\bar{l}}^{\theta}\}-\{\theta_{1},\Pi _{1}^{\theta}f_{\theta_{\bar{l}}}\}, 
\notag \\
&=&\{f(\theta_{\bar{l}},\lambda _{A}),\bar{\Pi}_{\bar{l}}^{\theta}\}-\{f(\theta_{\bar{l}
},\lambda _{A}),\Pi _{1}^{\theta}\}f_{\theta_{\bar{l}}},  \notag \\
&=&f_{\theta_{\bar{l}}}-f_{\theta_{\bar{l}}}=0.  
\end{eqnarray}
\begin{equation}
\{\theta_{\bar{l}},\Pi _{1}^{\theta}\}=\{\theta_{\bar{l}},\frac{\bar{\Pi}_{1}^{\lambda }}{
f_{\lambda _{1}}}\}=0.
\end{equation}
\begin{equation}
\{\theta_{\bar{l}},\Pi _{\bar{m}}^{\theta}\}=\{\theta_{\bar{l}},\bar{\Pi}_{\bar{m}}^{\theta}-\Pi
_{1}^{\theta}f_{\theta_{\bar{m}}}\}=\{\theta_{\bar{l}},\bar{\Pi}_{\bar{m}}^{\theta}\}=\delta _{
\bar{l}\bar{m}}.
\end{equation}
The previous calculations show that the algebra (\ref{commutationfree}) of
the standard MGT theory is recovered from the algebra (\ref
{commutation_constric2}) of the ENM. The transformation between both models
is given by (\ref{transformation2}).\ \ 
\subsection{The $\protect\phi _{\bar{A}}^{1}-(\theta_{j},\Pi _{j}^{\theta})$ sector}
The primary constraints are given by 
\begin{equation}
\phi _{\bar{A}}^{1}=\bar{\Pi}_{\bar{A}}^{\lambda }-\frac{\bar{\Pi}
_{1}^{\lambda }}{f_{\lambda _{1}}}f_{\lambda _{\bar{A}}}.
\end{equation}
Therefore 
\begin{eqnarray}
\{\phi _{\bar{A}}^{1},\theta_{1}\} &=&\{\phi _{\bar{A}}^{1},f(\theta_{\bar{l}},\lambda
_{A})\},  \notag \\
&=&\{\bar{\Pi}_{\bar{A}}^{\lambda },f(\theta_{\bar{l}},\lambda _{A})\}-\frac{
f_{\lambda _{\bar{A}}}}{f_{\lambda _{1}}}\bigg\{\bar{\Pi}_{1}^{\lambda
},\;f(\theta_{\bar{l}},\lambda _{A})\bigg\},  \notag \\
&=&-f_{\lambda _{\bar{A}}}+\frac{f_{\lambda _{\bar{A}}}}{f_{\lambda _{1}}}
f_{\lambda _{1}}=-f_{\lambda _{\bar{A}}}+f_{\lambda _{\bar{A}}}=0.
\end{eqnarray}
\begin{equation}
\{\phi _{\bar{A}}^{1},\theta_{\bar{l}}\}=\{\bar{\Pi}_{\bar{A}}^{\lambda }-\frac{
\bar{\Pi}_{1}^{\lambda }}{f_{\lambda _{1}}}f_{\lambda _{\bar{A}}},\theta_{\bar{l}
}\}=0.
\end{equation}
\begin{eqnarray}
\{\phi _{\bar{A}}^{1},\Pi _{1}^{\theta}\} &=&\bigg\{\phi _{\bar{A}}^{1},\frac{
\bar{\Pi}_{1}^{\lambda }}{f_{\lambda _{1}}}\bigg\},  \notag \\
&=&\bigg\{\bar{\Pi}_{\bar{A}}^{\lambda },\frac{\bar{\Pi}_{1}^{\lambda }}{
f_{\lambda _{1}}}\bigg\}-\bigg\{\frac{\bar{\Pi}_{1}^{\lambda }}{f_{\lambda
_{1}}}f_{\lambda _{\bar{A}}},\frac{\bar{\Pi}_{1}^{\lambda }}{f_{\lambda _{1}}
}\bigg\},  \notag \\
&=&\bigg\{\bar{\Pi}_{\bar{A}}^{\lambda },\frac{1}{f_{\lambda _{1}}}\bigg\}
\bar{\Pi}_{1}^{\lambda }-\frac{\bar{\Pi}_{1}^{\lambda }}{f_{\lambda _{1}}}
\bigg\{f_{\lambda _{\bar{A}}},\bar{\Pi}_{1}^{\lambda }\bigg\}\frac{1}{
f_{\lambda _{1}}},  \notag \\
&=&\frac{\bar{\Pi}_{1}^{\lambda }}{(f_{\lambda _{1}})^{2}}f_{\lambda
_{1}\lambda _{\bar{A}}}-\frac{\bar{\Pi}_{1}^{\lambda }}{(f_{\lambda
_{1}})^{2}}f_{\lambda _{\bar{A}}\lambda _{1}}=0.
\end{eqnarray}
\begin{eqnarray}
\{\phi _{\bar{A}}^{1},\Pi _{\bar{m}}^{\theta}\} &=&\{\phi _{\bar{A}}^{1},\bar{\Pi}
_{\bar{m}}^{\theta}-\Pi _{1}^{\theta}f_{\theta_{\bar{m}}}\},  \notag \\
&=&\{\phi _{\bar{A}}^{1},\bar{\Pi}_{\bar{m}}^{\theta}\}-\{\phi _{\bar{A}
}^{1},f_{\theta_{\bar{m}}}\}\Pi _{1}^{\theta},  \notag \\
&=&\bigg\{\bar{\Pi}_{\bar{A}}^{\lambda }-\frac{\bar{\Pi}_{1}^{\lambda }}{
f_{\lambda _{1}}}f_{\lambda _{\bar{A}}},\bar{\Pi}_{\bar{m}}^{\theta}\bigg\}-
\bigg\{\bar{\Pi}_{\bar{A}}^{\lambda }-\frac{\bar{\Pi}_{1}^{\lambda }}{
f_{\lambda _{1}}}f_{\lambda _{\bar{A}}},f_{\theta_{\bar{m}}}\bigg\}\Pi _{1}^{\theta}, 
\notag \\
&=&-\bar{\Pi}_{1}^{\lambda }\bigg\{\frac{f_{\lambda _{\bar{A}}}}{f_{\lambda
_{1}}},\bar{\Pi}_{\bar{m}}^{\theta}\bigg\}-
\bigg\{\bar{\Pi}_{\bar{A}}^{\lambda
},f_{\theta_{\bar{m}}}\bigg\}\Pi _{1}^{\theta}+\frac{f_{\lambda_{\bar{A}}}}{
f_{\lambda _{1}}}\bigg\{\bar{\Pi}_{1}^{\lambda },f_{\theta_{\bar{m}}}\bigg\}\Pi
_{1}^{\theta},  \notag \\
&=&-\bar{\Pi}_{1}^{\lambda }\bigg(\frac{f_{\lambda _{\bar{A}}\theta_{\bar{m}}}}{
f_{\lambda _{1}}}- \frac{f_{\lambda _{\bar{A}}}f_{\lambda _{1}\theta_{\bar{m}}}}{
(f_{\lambda _{1}})^{2}}\bigg)+f_{\theta_{\bar{m}}\lambda _{\bar{A}}}\Pi _{1}^{\theta}-
\frac{f_{\lambda _{\bar{A}}}}{f_{\lambda _{1}}}f_{\theta_{\bar{m}}\lambda
_{1}}\Pi _{1}^{\theta},  \notag \\
&=&-\Pi _{1}^{\theta}f_{\lambda _{\bar{A}}\theta_{\bar{m}}}+\Pi _{1}^{\theta}\frac{
f_{\lambda _{\bar{A}}}f_{\lambda _{1}\theta_{\bar{m}}}}{f_{\lambda _{1}}}+f_{\theta_{
\bar{m}}\lambda _{\bar{A}}}\Pi _{1}^{\theta}-\frac{f_{\lambda _{\bar{A}}}}{
f_{\lambda _{1}}}f_{\theta_{\bar{m}}\lambda _{1}}\Pi _{1}^{\theta}=0.\notag \\
\end{eqnarray}

\subsection{The $\protect\phi _{\bar{A}}^{1}-\protect\phi _{\bar{B}}^{1}$
sector}
\begin{eqnarray}
\{\phi _{\bar{A}}^{1},\phi _{\bar{B}}^{1}\} &=&\bigg\{\bar{\Pi}_{\bar{A}
}^{\lambda }-\frac{\bar{\Pi}_{1}^{\lambda }}{f_{\lambda _{1}}}f_{\lambda _{
\bar{A}}},\bar{\Pi}_{\bar{B}}^{\lambda }-\frac{\bar{\Pi}_{1}^{\lambda }}{
f_{\lambda _{1}}}f_{\lambda _{\bar{B}}}\bigg\},  \notag \\
&=&-\bigg\{\bar{\Pi}_{\bar{A}}^{\lambda },\frac{f_{\lambda _{\bar{B}
}}}{f_{\lambda _{1}}}\bigg\}\bar{\Pi}_{1}^{\lambda }-\bar{\Pi}
_{1}^{\lambda }\bigg\{\frac{f_{\lambda _{\bar{A}}}}{f_{\lambda _{1}}},\bar{
\Pi}_{\bar{B}}^{\lambda }\bigg\}+\bigg\{\frac{\bar{\Pi}_{1}^{\lambda
}}{f_{\lambda _{1}}}f_{\lambda _{\bar{A}}},\frac{\bar{\Pi}_{1}^{\lambda }}{
f_{\lambda _{1}}}f_{\lambda _{\bar{B}}}\bigg\},  \notag \\
&=&\frac{f_{\lambda _{\bar{B}}\lambda _{\bar{A}}}}{f_{\lambda _{1}}}
\bar{\Pi}_{1}^{\lambda }-\frac{f_{\lambda _{\bar{B}}}f_{\lambda
_{1}\lambda _{\bar{A}}}}{(f_{\lambda _{1}})^{2}}\bar{\Pi}_{1}^{\lambda }-
\frac{f_{\lambda _{\bar{A}}\lambda _{\bar{B}}}}{f_{\lambda _{1}}}
\bar{\Pi}_{1}^{\lambda }+\frac{f_{\lambda _{\bar{A}}}f_{\lambda _{1}\lambda
_{\bar{B}}}}{(f_{\lambda _{1}})^{2}}\bar{\Pi}_{1}^{\lambda }  \notag
\\
&&+\bar{\Pi}_{1}^{\lambda }\bigg\{\frac{f_{\lambda _{\bar{A}}}}{f_{\lambda
_{1}}},\bar{\Pi}_{1}^{\lambda }\bigg\}\frac{f_{\lambda _{\bar{B}}}}{
f_{\lambda _{1}}}+\frac{f_{\lambda _{\bar{A}}}}{f_{\lambda _{1}}}\bigg\{\bar{
\Pi}_{1}^{\lambda },\frac{f_{\lambda _{\bar{B}\;\;\;\;}}}{f_{\lambda _{1}}}
\bigg\}\bar{\Pi}_{1}^{\lambda },  \notag \\
&=&-\frac{f_{\lambda _{\bar{B}}}f_{\lambda _{1}\lambda _{\bar{A}}}}{
(f_{\lambda _{1}})^{2}}\bar{\Pi}_{1}^{\lambda }+\frac{f_{\lambda _{\bar{A}
}}f_{\lambda _{1}\lambda _{\bar{B}\;\;\;\;}}}{(f_{\lambda _{1}})^{2}}\bar{\Pi
}_{1}^{\lambda }  \notag \\
&&+\frac{\bar{\Pi}_{1}^{\lambda }}{f_{\lambda _{1}}}\bigg(\frac{f_{\lambda _{
\bar{B}}}f_{\lambda _{\bar{A}}\lambda _{1}}}{f_{\lambda _{1}}}-\frac{
f_{\lambda _{\bar{B}}}f_{\lambda _{\bar{A}}}f_{\lambda _{1}\lambda
_{1}}}{(f_{\lambda _{1}})^{2}}\bigg)-\frac{\bar{\Pi}_{1}^{\lambda }}{
f_{\lambda _{1}}}\bigg(\frac{f_{\lambda _{\bar{A}}}f_{\lambda _{\bar{B}
}\lambda _{1}}}{f_{\lambda _{1}}}-\frac{f_{\lambda _{\bar{A}
}}f_{\lambda _{\bar{B}}}f_{\lambda _{1}\lambda _{1}}}{(f_{\lambda
_{1}})^{2}}\bigg),  \notag \\
&=&-\frac{f_{\lambda _{\bar{B}}}f_{\lambda _{1}\lambda _{\bar{A}}}}{
(f_{\lambda _{1}})^{2}}\bar{\Pi}_{1}^{\lambda }+\frac{f_{\lambda _{\bar{A}
}}f_{\lambda _{1}\lambda _{\bar{B}}}}{(f_{\lambda _{1}})^{2}}\bar{\Pi
}_{1}^{\lambda }  \notag \\
&&+\frac{\bar{\Pi}_{1}^{\lambda }}{f_{\lambda _{1}}}\bigg(\frac{f_{\lambda _{
\bar{B}}}f_{\lambda _{\bar{A}}\lambda _{1}}}{f_{\lambda _{1}}}-\frac{
f_{\lambda _{\bar{A}}}f_{\lambda _{\bar{B}}\lambda _{1}}}{f_{\lambda
_{1}}}\bigg)=0.
\end{eqnarray}

\subsection{The $\protect\phi _{\bar{A}}^{1}-\protect\phi _{\bar{B}
}^{2}$ sector}
\begin{equation}
\phi _{\bar{A}}^{1}=\bar{\Pi}_{\bar{A}}^{\lambda }-\frac{\bar{\Pi}
_{1}^{\lambda }}{f_{\lambda _{1}}}f_{\lambda _{\bar{A}}},\quad \quad \quad
\phi _{\bar{B}}^{2}=f_{\lambda _{\bar{B}}}-\frac{f_{\lambda
_{1}}}{G_{1}}G_{\bar{B}},
\end{equation}
\begin{equation}
\{\phi _{\bar{A}}^{1},\phi _{\bar{B}}^{2}\}=\{\phi _{\bar{A}
}^{1},f_{\lambda _{\bar{B}}}\}-\frac{G_{\bar{B}}}{G_{1}}
\{\phi _{\bar{A}}^{1},f_{\lambda _{1}}\},
\end{equation}
where we have used that $\{\phi _{\bar{A}}^{1},G_{m}(\theta_{l},\Pi _{m}^{\theta})\}=0$,
 because we have proved that $\{\phi _{\bar{A}}^{1},\theta_{l}\}=\{\phi _{\bar{A}
}^{1},\Pi _{m}^{\theta}\}=0$. Therefore 
\begin{eqnarray}
\label{matrizx1}
\{\phi _{\bar{A}}^{1},\phi _{\bar{B}}^{2}\} &=&\{\bar{\Pi}_{\bar{A}
}^{\lambda },f_{\lambda _{\bar{B}}}\}-\frac{f_{\lambda _{\bar{A}}}}{
f_{\lambda _{1}}}\bigg\{\bar{\Pi}_{1}^{\lambda },f_{\lambda _{\bar{B}
}}\bigg\}-\frac{G_{\bar{B}}}{G_{1}}\bigg(\{\bar{\Pi}_{\bar{A}
}^{\lambda },\;f_{\lambda _{1}}\}-\frac{f_{\lambda _{\bar{A}}}}{f_{\lambda
_{1}}}\bigg\{\bar{\Pi}_{1}^{\lambda },\;f_{\lambda _{1}}\bigg\}\bigg), \notag \\
&=&-f_{\lambda _{\bar{B}}\lambda _{\bar{A}}}+\frac{f_{\lambda _{\bar{A}}}}{
f_{\lambda _{1}}}f_{\lambda _{\bar{B}}\lambda _{1}}-\frac{G_{\bar{B}}
}{G_{1}}\bigg(f_{\lambda _{1}\lambda _{\bar{A}}}-\frac{f_{\lambda _{\bar{A}}}
}{f_{\lambda _{1}}}f_{\lambda _{1}\lambda _{1}}\bigg),  \notag \\
&=&-f_{\lambda _{\bar{B}}\lambda _{\bar{A}}}+\frac{f_{\lambda _{\bar{A}}}}{
f_{\lambda _{1}}}f_{\lambda _{\bar{B}}\lambda _{1}}-\frac{f_{\lambda _{\bar{B
}\;\;\;\;}}}{f_{\lambda _{1}}}\bigg(f_{\lambda _{1}\lambda _{\bar{A}}}-\frac{
f_{\lambda _{\bar{A}}}}{f_{\lambda _{1}}}f_{\lambda _{1}\lambda _{1}}\bigg),
\end{eqnarray}
where in the last line we have employed the constraint $\phi _{\bar{B}}^{2}=0$ to obtain 
\begin{equation}
\frac{G_{\bar{B}}}{G_{1}}=\frac{f_{\lambda _{\bar{B}}}}{
f_{\lambda _{1}}}.
\end{equation}
The invertibility of the matrix $T_{\bar{A}\bar{B}}=\{\phi _{\bar{A}
}^{1},\phi _{\bar{B}}^{2}\}$ depends of the function $f=f(\theta_{\bar{l}
},\lambda _{m})$. A direct calculation shows that if we take 
\begin{equation}
f=f(\Psi ),\quad \quad \quad \quad \Psi =\sum_{i}\frac{\tilde{X}_{i}^{2}}{2},
\label{particularNM}
\end{equation}
where $\tilde{X}_{i}$ denotes the variables  $\theta_{\bar{l}}$ and $\lambda _{A}$, 
the matrix $T_{\bar{A}\bar{B}}=\{\phi _{\bar{A}}^{1},\phi _{\bar{B}}^{2}\}$
becomes 
\begin{equation}
T_{\bar{A}\bar{B}}=\{\phi _{\bar{A}}^{1},\phi _{\bar{B}}^{2}\}=f^{\prime
}\times \bigg(\delta _{\bar{A}\bar{B}}+\frac{\lambda _{\bar{B}\;}\lambda _{
\bar{A}}}{\lambda _{1}^{2}}\bigg),\;\;\;\;f^{\prime }=\frac{df}{d\Psi },
\end{equation}
which is invertible with  
\begin{equation}
(T^{-1})_{\bar{A}\bar{B}}=\frac{1}{f^{\prime }}\times \bigg(\delta _{
\bar{A}\bar{B}}-\frac{\lambda _{\bar{B}}\lambda _{\bar{A}}}{\lambda _{c}^{2}}
\bigg),\quad \quad \quad c=1,2,...,K.
\end{equation}
The function (\ref{particularNM}) is a generalization of the constraint $A_\mu A^\mu=n^2M^2$ 
used to define the abelian Nambu model (\ref{NANMabeliano}), which in this case 
would be written as
\begin{equation}
A_1=\sqrt{A_0^2-n^2M^2-A_jA_j},\quad\quad\quad j=1,2,
\end{equation}
where we have made the identifications
\begin{equation}
\theta_1=\theta_1(\theta_{\bar{l}},\lambda_A)=f(\theta_{\bar{l}},\lambda_A) \, \rightarrow \, A_1=A_1(A_j,A_0)=\sqrt{A_0^2-n^2M^2-A_jA_j},\, j=1,2.
\end{equation} 
In this case, the constraint $F=F(\theta_l,\lambda_A)=0 \,$ is subjected to the following conditions: (i) It is possible to solve for $\theta_1$ and (ii) the function $\theta_1=f(\theta_{\bar{l}},\lambda_A)$ is such that the matrix defined by Eq. (\ref{matrizx1}) is invertible.

\section{The algebra resulting when solving for the coordinate $\protect\lambda _{1}$
in the ENM constraint}
\label{Apendice2}
In this situation the calculation is direct since the canonical variables of
the ENM and those of the MGT coincide.
\subsection{The $\protect\phi^1_{\bar{m}}-(\theta_i,\Pi^\theta_i)$ sector}
In this case the primary constraints are $\phi _{\bar{b}}^{1}=\Pi _{\bar{b}}^{\lambda }$ and the canonical
algebra of the ENM gives 
\begin{equation}
\{\phi _{\bar{b}}^{1},\theta_{i}\}=\{\phi _{\bar{b}}^{1},\Pi _{i}^{\theta}\}=0.
\end{equation}
\subsection{The $\protect\phi^1_{\bar{m}}-\protect\phi^1_{\bar{m}}$ sector}
\begin{equation}
\{\phi^1_{\bar{a}},\phi^1_{\bar{b}}\}=\{\Pi^\lambda_{\bar{a}},\Pi^\lambda_{
\bar{b}}\}=0.
\end{equation}
\subsection{The $\protect\phi^1_{\bar{m}}-\protect\phi^2_{\bar{m}}$ sector}
\begin{equation}
\phi _{\bar{b}}^{1}=\Pi _{\bar{b}}^{\lambda },\quad \quad \quad \phi _{\bar{b
}}^{2}=g_{\lambda _{\bar{b}}}+\frac{G_{\bar{b}}}{G_{1}}.
\end{equation}
\begin{equation}
X_{\bar{a}\bar{b}}=\{\phi _{\bar{a}}^{1},\phi _{\bar{b}}^{2}\}=\bigg\{\Pi _{
\bar{a}}^{\lambda },g_{\lambda _{\bar{b}}}+\frac{G_{\bar{b}}}{G_{1}}\bigg\}
=g_{\lambda _{\bar{b}}\lambda _{\bar{a}}}.
\label{matrix22}
\end{equation}
Again, we can take as an example 
\begin{equation}
g=g(\Psi ),\quad \quad \quad \quad \Psi =\sum_{i}\frac{\tilde{X}_{i}^{2}}{2},
\label{particularNM2}
\end{equation}
where $\tilde{X}_{i}$ denotes the variables  $\theta_{l}$ and $\lambda _{\bar{m}}$.
The matrix $X_{\bar{l}\bar{m}}=\{\phi _{\bar{l}}^{1},\phi _{\bar{m}}^{2}\}$
becomes 
\begin{equation}
X_{\bar{l}\bar{m}}=\{\phi _{\bar{l}}^{1},\phi _{\bar{m}}^{2}\}=g^{\prime
}\times \bigg(\delta _{\bar{l}\bar{m}}+\frac{\lambda _{\bar{m}}\lambda _{
\bar{l}}}{g^{\prime }}g^{\prime \prime }\bigg),\quad\quad\quad g^\prime= \frac{dg}{d\Psi},
\end{equation}
which is invertible 
\begin{equation}
(X^{-1})_{\bar{l}\bar{m}}=\frac{1}{g^{\prime }}\times \bigg(\delta _{\bar{l}
\bar{m}}-\frac{g^{\prime \prime }}{(g^{\prime }+\lambda _{\bar{c}}\lambda _{
\bar{c}}g^{\prime \prime })}\lambda _{\bar{m}}\lambda _{\bar{l}}\bigg).
\end{equation}
The function (\ref{particularNM2}) is a generalization of the constraint $A_\mu A^\mu=n^2M^2$ 
used to define the abelian Nambu model (\ref{NANMabeliano}), which in this case 
would be written as
\begin{equation}
A_0=\sqrt{n^2M^2+A_iA_i},\quad i=1,2,3,
\end{equation}
after making the identification
\begin{equation}
\lambda_1=\lambda_1(\theta_l,\lambda_{\bar{A}})=g(\theta_l,\lambda_{\bar{A}})\rightarrow A_0=A_0(A_i)=\sqrt{n^2M^2+A_iA_i},\quad\quad\quad i=1,2,3.
\end{equation}
In this case, the requirements upon the constraint $F=F(\theta_l,\lambda_A)=0$ are: (i) It is possible to solve for $\lambda_1$ and (ii) the function $\lambda_1=g(\theta_l,\lambda_{\bar{A}})$ is such that the matrix defined by Eq. (\ref{matrix22}) is invertible.

\end{document}